\documentclass[manuscript]{acmart}

\setcopyright{none}
\renewcommand\footnotetextcopyrightpermission[1]{}
\usepackage{multirow}
\usepackage{graphicx}
\usepackage{xcolor}
\usepackage{algorithmic}
\usepackage{url}
\usepackage{enumitem}
\usepackage{array}
\usepackage{textcomp}
\usepackage{listings}
\usepackage{float}
\usepackage{makecell}
\usepackage{cancel}
\usepackage{color,soul}
\usepackage[font=footnotesize]{subcaption}
\usepackage{framed}
\usepackage{rotating}

\usepackage{wrapfig}

\def\RQthree{How do users attempt to mitigate failures, and how effective are these strategies?}
\def\RQfour{How do users perceive and adapt to imperfect LLM responses when performing SE tasks?}

\def\RQone{Under what conditions do LLMs fail to generate helpful answers?}
\def\RQtwo{What are the underlying reasons for the observed failures?}

\begin{document}

\title{"Should I Give Up Now?" Investigating LLM Pitfalls in Software Engineering}

\author{Jiessie Tie}
\affiliation{
    \institution{University of Toronto}
    \city{Toronto}
    \country{Canada}
}
\email{jiessie.tie@mail.utoronto.ca}

\author{Bingsheng Yao}
\affiliation{
    \institution{Northeastern University}
    \city{Boston}
    \country{USA}
}
\email{b.yao@northeastern.edu}

\author{Tianshi Li}
\affiliation{
    \institution{Northeastern University}
    \city{Boston}
    \country{USA}
}
\email{tia.li@northeastern.edu}

\author{Hongbo Fang}
\affiliation{
    \institution{University of Chicago}
    \city{Chicago}
    \country{USA}
}
\email{fanghongdoublebo@gmail.com}

\author{Syed Ishtiaque Ahmed}
\affiliation{
    \institution{University of Toronto}
    \city{Toronto}
    \country{Canada}
}
\email{ishtiaque@cs.toronto.edu}

\author{Dakuo Wang}
\affiliation{
    \institution{Northeastern University}
    \city{Boston}
    \country{USA}
}
\email{d.wang@northeastern.edu}

\author{Shurui Zhou}
\affiliation{
    \institution{University of Toronto}
    \city{Toronto}
    \country{Canada}
}
\email{shurui.zhou@utoronto.ca}

\begin{abstract}
Software engineers are increasingly incorporating AI assistants into their workflows to enhance productivity and alleviate cognitive load. 
However, experiences with large language models (LLMs) such as ChatGPT vary widely. 
While some engineers find them useful, others deem them counterproductive due to inaccuracies in their responses. Researchers have also observed that ChatGPT often provides incorrect information. Given these limitations, it is crucial to determine how to effectively integrate LLMs into software engineering (SE) workflow.
 Analyzing data from 26 participants in a complex web development task, we identified nine failure types categorized into incorrect or incomplete responses, cognitive overload, and context loss. Users attempted to mitigate these issues through scaffolding, prompt clarification, and debugging. However, 17 participants ultimately chose to abandon ChatGPT due to persistent failures. 
Our quantitative analysis revealed that unhelpful responses increased the likelihood of abandonment by a factor of 11, while each additional prompt reduced abandonment probability by 17\%. 
This study advances the understanding of human-AI interaction in SE tasks and outlines directions for future research and tooling support.
\end{abstract}

\setcopyright{cc}
\setcctype{by}
\acmJournal{TOSEM}
\acmYear{2026} \acmVolume{1} \acmNumber{1} \acmArticle{}
\acmMonth{1} \acmDOI{10.1145/3801972}



\received{June 7 2025}
\received[revised]{December 10 2025}
\received[accepted]{March 7 2025}

\maketitle

\section{Introduction and Background}
The rise of LLM-powered assistants (e.g., ChatGPT~\cite{brown2020language}, CoPilot~\cite{chen2021evaluating}, 
CursorAI ~\cite{Cursor}
)

has introduced innovative methods for integrating AI into traditionally human tasks. ChatGPT enhances productivity~\cite{schmidhuber_2021}, improves work quality~\cite{sridhara_2023}, and is widely accessible to students and professionals in computer science and SE~\cite{okonkwo_2021}. It supports learning, task completion~\cite{kazemitabaar_2023, tian_2023, mock2024generative}, and automates repetitive tasks to reduce cognitive load~\cite{sridhara_2023, schmidhuber_2021, Guo_2023}.

However, researchers also have identified limitations, including potential decreased cognitive skills~\cite{heersmink2024use} and over-reliant behaviors among novices ~\cite{kazemitabaar2023novices}.
Although prompt engineering has been explored~\cite{dang_2022}, users rarely apply these techniques effectively~\cite{khurana2024and}. 
Additionally, LLM responses vary across iterations, and the effectiveness of prompting techniques may decline in newer models, sometimes leading to worse performance~\cite{wang2024advanced}. This variability necessitates continuous refinement of prompt engineering strategies and can frustrate users who must repeatedly adjust their approaches~\cite{muller2015stuck}.
While prior work has examined LLM capabilities and prompting strategies under various conditions, it has less explicitly focused on how users interact with LLMs when structured prompting support is limited. This motivates closer examination of realistic usage conditions.

 Unlike conventional SE tools, LLMs function much more like pair-programmers: 
they generate evolving, context-dependent outputs that require developers to continually reinterpret, validate, and adapt their next steps \cite{yetistiren2022assessing, tang2024study}. This multi-turn, co-adaptive nature introduces breakdowns that differ fundamentally from traditional usability issues. Developers not only consume information but must monitor, debug, and integrate model outputs into their workflow, often without guarantees of consistency or correctness.
Although prior CHI and SE research has examined tool failures, usability breakdowns, and adoption barriers, these frameworks assume stable, deterministic tools and therefore do not account for the multi-turn, probabilistic, and co-adaptive nature of LLM interactions. As a result, existing theories cannot fully explain how LLM-specific failures accumulate across turns or why users abandon LLM support during task execution.

 These properties make the study of LLM failures and abandonment a distinct and increasingly important research area. Traditional SE and HCI tool-adoption literature generally assumes that tools behave deterministically and provide stable and predictable feedback \cite{harman2012role, dooner1975application, reiss2017framework}. LLMs, in contrast, may hallucinate, omit important steps, lose context across turns, or produce misleading code \cite{nguyen2022empirical, imai2022github}. These issues accumulate over time and influence the developer’s reasoning process in ways that prior models do not account for \cite{li2025always, kabir_2024, zhong_2024}. When such failures persist, developers may abandon the tool in the middle of a task. This type of abandonment has not been theorized or measured in earlier work on tool adoption or programming assistance. As a result, studying how failures occur, how users attempt to mitigate them, and why they eventually give up is essential for both human-AI interaction theory and the practical integration of LLMs into SE workflows.

Understanding how users navigate LLM failures and adapt their interactions during multi-turn tasks is an increasingly important research focus, particularly because these dynamics differ from those documented in studies of traditional SE or HCI tools.
Prior studies have explored LLM usability and identified common challenges ~\cite{khurana2024and, chen2021evaluating, denny_2024, jalil_2023,
balhorn2023does}. However, much of this research relies on large-scale surveys\cite{liang2024large} or narrowly scoped use cases~\cite{choudhuri_2023}, which may not fully capture the complexities of real-time human-LLM interactions. 
A more detailed investigation into the fine-grained interactions between users and LLMs can offer a clearer understanding of how failures occur and how users attempt to navigate them.
To address this gap, we conducted observational studies in which participants completed a multi-step web development task using  ChatGPT, due to its accessibility when working with unfamiliar tasks through its intuitive interface and SOTA code-solving ability. 
Our goal is to understand the realistic, low-scaffolding usage rather than assessing the upper-bound model capabilities with expert-optimized prompting.
Our study aimed to answer the following research questions (RQs):

\begin{itemize}[leftmargin=1em]
     \item \textbf{RQ1:} \RQone
    \item \textbf{RQ2:}  \RQtwo
    \item \textbf{RQ3:} \RQthree 
    \item  \textbf{RQ4:}  \RQfour
    \item \textbf{RQ5:}  What user characteristics and interaction patterns are associated with abandoning LLM assistance during tasks?
\end{itemize}

 \noindent
Tab.~\ref{tab:rq_mapping} provides a consolidated mapping between each research question and its underlying construct, operational measures, and corresponding data sources. This mapping makes explicit how our qualitative and quantitative methods are grounded in well-defined constructs. RQ1–RQ4 investigate constructs such as LLM helpfulness, failure causes, and user mitigation or adaptation, which we operationalize through systematic coding of interaction logs, code outputs, and interview reflections. RQ5 extends this framework by applying a quantitative model to examine how operationalized factors such as helpfulness, task performance, prompt frequency, and coding experience relate to abandonment behavior. 
With this conceptual and methodological structure established, we next describe our study design and participant sample. 

\begin{table*}[t]
\centering
\footnotesize
\setlength{\tabcolsep}{4pt}
\renewcommand{\arraystretch}{1.15}
\caption{Mapping of research questions to their corresponding constructs, operational measures, and data sources. 
This table clarifies how each construct in our study is defined and measured, and how qualitative and quantitative data sources support the analysis for each research question.}
\label{tab:rq_mapping}
\begin{tabular}{@{}p{0.5cm}p{2cm}p{7cm}p{3cm}@{}}
\toprule
\textbf{RQ} & \textbf{Constructs} & \textbf{Operationalization} & \textbf{Data Sources}\\
\midrule
\textbf{RQ1} & LLM helpfulness, correctness, completeness & Manual coding of LLM outputs into helpful, partially helpful, or unhelpful categories; task completeness rubric & ChatGPT response, implemented code output\\
\midrule
\textbf{RQ2} & Failure causes & Manual coding of user- and LLM-related causes & ChatGPT response, code-level discrepancies\\
\midrule
\textbf{RQ3} & User mitigation strategies & Manual coding of user responses following failed LLM outputs & ChatGPT response, user actions from screen recordings, and interview comments
\\
\midrule
\textbf{RQ4} & User adaptation and perception & Manual coding of post-task reflections & Semi-structured interview transcripts\\
\midrule
\textbf{RQ5} & Abandonment behavior & 
Logistic regression model predicting \textit{Abandon\_ChatGPT} from helpfulness, task performance, prompt frequency, and coding experience, and interaction-level indicators of disengagement (e.g., decreasing prompt rate, last-turn timestamp, cessation of LLM querying)
& 
LLM prompt histories, interaction timestamps, user background characteristics
\\
\bottomrule
\end{tabular}
\end{table*}

We initially conducted the study with 21 students, followed by a small-scale replication with five professional software developers (SDEs)  to examine the analytic consistency and transferability of interaction patterns across experience levels.
Across both groups, we observed analytically consistent failure patterns and mitigation behaviors: despite substantial differences in programming expertise, students and SDEs engaged in similar interaction practices and encountered comparable challenges when using ChatGPT for task completion.

Given the rapid evolution of LLMs and the obsolescence of earlier model versions during the revision cycle, we additionally replicated the same experimental protocol with new participants using GPT-5.1. This extension was intended to ensure that our findings remain relevant to current state-of-the-art systems rather than to evaluate model performance per se. The GPT-5.1 replication revealed the same categories of failures and mitigation strategies as earlier models. Although GPT-5.1 improved performance on some one-shot code generation tasks, the same underlying interactional breakdowns persisted during multi-step task execution.

Overall, we identified 92 failure cases (58\%) across nine distinct types, which can be grouped into three categories: `incomplete or incorrect responses', `cognitive overload and information mismanagement', and `lack of context retention and redundant effort' (RQ1).
Our analysis uncovered 12 underlying causes (RQ2), stemming from both user miscommunication  (e.g., missing details from skimming) and ChatGPT's limitations  (e.g., ignoring users' expertise, ignoring instructions). Additionally, we identified seven mitigation strategies employed by participants (RQ3). 

From our post-study reflection, we assessed user perceptions of ChatGPT’s effectiveness (RQ4).  While ChatGPT reduces coding efforts, prompting can be equally tedious. SDEs frequently using ChatGPT for rapid code generation and summarization of unfamiliar tasks often struggle to grasp the underlying structure, hindering learning. 

In addition, our hypothesis testing results (RQ5) indicate that unhelpful responses significantly increased the likelihood of ChatGPT abandonment (odds ratio = 11). Conversely, a higher number of prompting iterations and greater coding experience (e.g., SDEs compared to students) were associated with lower abandonment rates.
In summary, the primary contributions of this paper are:

\begin{itemize}[leftmargin=1em]
    \item We use fine-grained observations with objective user behavior measures, presenting a dataset of interaction histories with ChatGPT for complex SE tasks, which could serve as a valuable resource for future research endeavors.
    \item We conduct a comprehensive analysis of human-LLM interactions, identifying nine distinct failure types, along with 12 underlying causes.
    \item We provide an overview of the existing mitigation strategies that users employ to address failure cases and propose potential research and tool development opportunities.
    \item We present novel findings supported by statistical evidence, revealing how user behaviors influence LLM effectiveness, extending prior research beyond model limitations in SE tasks.

\end{itemize}

\section{Related Work}
Given the rapid expansion of LLM research, this section provides an overview of current studies on the use and evaluation of LLMs in SE. 
We highlight representative works that share methodological or thematic similarities with our study, particularly those analyzing ChatGPT interaction failures in SE tasks. 
A more comprehensive list of related papers is available in the replication package~\cite{anonymous_2024_13179522}.

\subsection{LLM for SE} 

Developers are increasingly adopting conversational agents, such as ChatGPT, Copilot and Gemini, to support their software development processes \cite{li2025always}, reflecting a broader shift toward AI-assisted coding that enhances both cognitive and productivity outcomes. 
These tools offer opportunities for more efficient debugging, context-aware explanations, and smoother workflow integration \cite{sadat2024generative, mock2024generative}.
Specifically, studies identify cognitive support benefits, including improved code comprehension~\cite{nam_2024, yuan2023evaluating}, productivity~\cite{xu_2022}, learning integration~\cite{daun2023chatgpt, yilmaz_2023, kuhail_2023}, and explanation-based reasoning that aids conceptual understanding~\cite{kazemitabaar_2024, nam_2024}.
In parallel, research also highlights tool-level support, such as IDE support~\cite{warner2017codepilot}, enhanced productivity \cite{xu_2022}, and applications in programming education \cite{ouh2023chatgpt, Guo_2023}.

However, while prior studies on conversational agents such as ChatGPT and Copilot demonstrate benefits for code comprehension, productivity, and learning \cite{khojah2024beyond, chouchenHowWareDevelopers2024, meem2025software, sagdic2024taxonomy} they largely focus on short, individual interactions, leaving open how such tools support complex or collaborative programming contexts. Prior studies that evaluate LLMs on short or isolated tasks do not capture the multi-step reasoning processes where failures accumulate, and abandonment decisions emerge.
These issues become more pronounced in complex programming tasks, where users must continuously interpret and adapt to evolving model behavior.

\subsubsection{Understanding User-LLM Factors for Task Completion}
A growing body of work focuses on improving LLM output from the system, and the designer's perspective ~\cite{coignion2024performance, garg2025rapgen, vaithilingam_2022, qiu2024efficient}. Techniques, such as retrieval-augmented generation (RAG)~\cite{arslan2024survey}, context engineering~\cite{mei2025survey}, instruction-tuning~\cite{peng2023instruction} and fine-tuning\cite{ouyang2022training}, have proved to be helpful through benchmark validation studies for automating issue solving~\cite{ceurstemont2025automating, ma2025alibaba}.
While these advances improve model capabilities, they primarily modify the model architecture or training process rather than the user interaction process.
However, from the user's end, there are fewer ways to interact with LLM systems, especially for realistic interactions for ill-structured, low-scaffolded tasks~\cite{sergeyuk2026human}. 

In practice, most user-LLM interactions occur through prompts, by users compiling written questions and evaluating and implementing LLM-generated answers. This interaction paradigm places substantial responsibility on the users to translate their goals into structured prompts and to understand and evaluate LLM outputs based on their knowledge level with the task. 

One user-side strategy that has become popularised to support user interactions without updating model designs is prompt engineering~\cite{white2023, giray2023prompt, liu2023pre}.
A number of prior works suggest that prompt engineering can improve LLM output~\cite{ye2025neural, white2023, liu2023pre, dang_2022}. 
There are different types of prompt engineering, to varying levels of effectiveness. Few-shot prompting has become widespread, where users optimize the quality of LLM output based on effective task-breakdown for task completion~\cite{brown2020language}.

However, applying prompt engineering in practice is non-trivial. 
In one study done by Zamfirescu-Pereira et al. \cite{zamfirescu-pereira_2023} show that non-AI experts struggle to design effective prompts. 
Similarly, Kruse et al. \cite{kruse2024can} find that professional and student developers were unaware of how to apply prompt-engineering techniques. These findings suggest that knowledge of prompting strategies does not reliably translate into effective real-world use, particularly during unaided development.

Moreover, the marginal utility of prompt engineering may be smaller than commonly assumed. 
Khojah et al. \cite{khojah2025impact} find that GPT is not very sensitive to prompt engineering. 
Additionally, other work suggests that newer LLMs are increasingly robust to prompt engineering~\cite{wang2024advanced}, further reducing the guarantee that more elaborate prompting will substantially improve output quality. For users, investing time and cognitive effort into learning prompt engineering may therefore not consistently yield proportional performance gains.

Context engineering is another way of improving model performance, but can be resource consuming, and not very suitable for human interactions \cite{mei2025survey}. 
Similarly, providing external resources to users in the form of model tutorials assumes that users proactively seek out and internalize such resources. But these are not intuitive for users unless they look for these resources themselves, which may not always hold.

As LLMs become increasingly accessible, non-experts are beginning to use them for ill-defined and complex tasks that previously required substantial domain expertise. The growing popularity of ``vibe-coding,'' where users rely on intuition rather than structured prompting frameworks~\cite{fawzy2025vibe}, reflects this shift toward informal, opportunistic interaction styles. Rather than assuming idealized prompt optimization, our study therefore examines how users actually interact with LLMs during unaided task completion, including how breakdowns emerge and how mitigation strategies are (or are not) developed and used in practice.

\subsubsection{Existing Challenges for LLMs in Software Development}
 Despite the benefits of LLMs, existing research highlights persistent technical, human, and perceptual challenges that limit LLMs' effectiveness in real-world programming contexts.

Technical limitations are often most salient during use, and constrain how reliably LLMs can reason about and generate code~\cite{fan_2023, tian_2023, kabir_2024, jimenez_2023}.
Prior work evaluating LLMs' autonomous ability on well-defined, single-step coding problems \cite{nguyen2022empirical, coignion2024performance} shows that models frequently generate incorrect or inconsistent code and require multiple follow-up prompts or human edits.
Efforts to benchmark and fine-tune code-focused models \cite{chen2021evaluating, liu2024your, yetistiren2022assessing} especially demonstrate the limitations in answer consistency and reasoning abilities, finding that LLMs can require several follow-up prompts to improve solutions. Studies on autonomous agents' performance on real-world SWE tasks find that LLMs still require much oversight as answers are still often incorrect ~\cite{sridhara_2023, surameery_2023, liu2025empirical}. 

Human oversight remains critical even when LLMs perform autonomous SE tasks, particularly because real-world programming problems are often ill-defined and lack clear solutions.
Autonomous agent evaluations show that human intervention is often necessary for ensuring correctness and functionality \cite{sridhara_2023, surameery_2023, liu2025empirical}.
Novel tasks also introduce challenges, such as difficulties in articulating successful prompts, resulting in suboptimal responses~\cite{zamfirescu-pereira_2023}, decreased user confidence~\cite{amoozadeh2024trust}, and reduced adoption of AI tools~\cite{farah2022impersonating}. 
Even professionals can face limitations outside of model outputs, with findings suggesting that failures can deter users from continuing with LLMs, even with perceived benefits to productivity ~\cite{weisz2025examining, khojah2024beyond}. 

Finally, user perception and adoption barriers further limit the sustained use of LLMs.
Researchers comparing ChatGPT responses to Stack Overflow questions found high rates of misinformation and factual errors, raising concerns about reliability \cite{kabir_2024, zhong_2024}. 
Large-scale surveys of AI-assisted programming similarly report mixed perceptions, where developers preferred AI tools for code generation abilities, but reported abandonment due to inconsistent model behavior \cite{jiang2024survey, jimenez_2023, fan_2023, liang2024large, nascimento2023comparing}. 
While some studies suggest that LLMs can provide expert-level responses to engineering questions \cite{balhorn2023does}, others show limited productivity gains even on basic programming tasks \cite{xu_2022, ouyang2025empirical, danyaro2025llm}. 
 Together, these findings highlight ongoing challenges in reliability and user trust, motivating the need to better understand how developers collaborate with LLMs in complex, real-world contexts.

Taken together, these limitations indicate that LLM failures often unfold across multiple interaction turns rather than in isolated instances. These accumulated breakdowns may influence whether users continue or abandon the tool, which highlights the importance of studying such dynamics in depth.

 Our work contributes a real-time observational study that links developer interaction behavior under realistic, low-scaffolding conditions to LLM output quality, revealing how interaction patterns influence the quality and reliability of generated code. 
Our study looks at the real-world  applications of ChatGPT for SE in an empirical setting, without optimising for prompt-engineering techniques.
To capture detailed instances of meaningful exchange between developers and LLMs, we categorize tasks based on: \textit{simple}, direct solution tasks and \textit{complex}, multi-step tasks.
    Simple programming tasks, such as writing a short function or fixing a syntax error, typically have clear goals, well-defined inputs and outputs, and limited uncertainty in the solution path.  \cite{barke_2023, nascimento2023comparing, wang2024rocks}
    In contrast, complex tasks can involve larger codebase or integrating unfamiliar frameworks that require multiple parts, larger contexts, and iterative problem decomposition. 
Prior work shows that LLM effectiveness varies sharply across these dimensions: while models perform reliably on short, well-specified prompts, they struggle with long-horizon reasoning, ambiguous requirements, and dependency management \cite{meem2025software, chouchenHowWareDevelopers2024}. 
Tab.~\ref{tab:related_works} summarizes key studies on evaluating LLMs for SE tasks, considering expertise levels, task complexity (simple vs. complex), and evaluation methods. 

Our study extends prior work with a fine-grained analysis of failure cases in human-LLM interactions for multi-component web application development. 
The detailed data granularity also enables both qualitative examination of interactional breakdowns and quantitative investigation of user characteristics and interaction patterns associated with abandoning LLM assistance during tasks.

\begin{table*}[!htbp]
\centering
\footnotesize
\caption{Comparison of Related Work and Our Approach}
\label{tab:related_works}
\setlength{\tabcolsep}{3pt}
\begin{tabular}{@{}p{0.8cm}p{0.8cm}p{1.6cm}p{2.5cm}p{2.5cm}p{5cm}@{}}
\toprule
\textbf{Ref.} & \textbf{Simple Task?} & \textbf{User Interactions?} & \textbf{Method (\#Observation Points)} & \textbf{Measurement} & \textbf{Distinction From Our Work}\\
\midrule
\cite{liang2024large} & Yes & User Self-Report & Survey (410) & Usability, Perceptions & We evaluate tasks experimentally with multiple dimensions.\\
\midrule
\cite{barke_2023} & Yes & Yes & Observational (20) & Completion time & We provide quantitative and qualitative analyses of multi-step tasks, with a focus on task failures.\\
\midrule
\cite{vaithilingam_2022} & Yes & Yes & Within-subj. (24) & Completion, Failures, Perceptions & We examine user behavior across broader programming scenarios.\\
\midrule
\cite{nascimento2023comparing} & Yes & No & Leetcode & Completion & Our approach captures detailed user–tool interactions beyond basic metrics.\\
\midrule
\cite{wang2024rocks} & Yes & Yes & Between-subj. (109) & Completion, observations & We explicitly study multi-step user decision-making processes.\\
\midrule
\cite{rasnayaka2024empirical} & No & User Self-Report & Observational (214 novices) & Completion, perceptions & We analyze user interactions with ChatGPT for unfamiliar tasks.\\
\midrule
\cite{choudhuri_2023} & No & User Self-Report & Between-subj. (22 novices) & Productivity, self-efficacy & We explore debugging behavior and propose strategies to mitigate adverse user behaviors.\\
\midrule
\cite{prather_2024} & No & Yes & Within-subj. (19 novices) & Perception & We extend evaluations to realistic programming scenarios beyond perception.\\
\midrule
\cite{mock2024generative} & No & No & Case Study (5) & Correctness, perception & Our evaluation framework is designed to be comprehensive and generalizable.\\
\midrule
\cite{liukko2024chatgpt} & No & No & Case Study (7) & Completion, observations & We explicitly consider error analysis and provide mitigation recommendations.\\
\midrule
\cite{ulfsnes2024transforming} & No & Interviews & Interview (13) & Perception & We focus specifically on SE-related challenges.\\
\midrule
\cite{khojah2024beyond} & No & Yes & Observational (24 professional developers) & Usage patterns, task purpose, perceived usefulness, trust & Examines ChatGPT use in real-world SE tasks; our study extends this by focusing on failures and abandonment during task completion.\\
\midrule
\cite{weisz2025examining} & No & User Self-Report & Field study & Productivity metrics, self-reported experience, qualitative interviews & Investigates AI assistant adoption in enterprise; our study complements by analyzing real-time interactions and identifying failure causes.\\
\midrule
\cite{chouchenHowWareDevelopers2024} & No & Yes & Mixed-method (243 PRs with ChatGPT prompts vs. 384 without) & Quantitative analysis of PR size, closure rate, and qualitative taxonomy of prompt topics & Analyzes large-scale PR data to characterize ChatGPT use in code review and maintenance; our study focuses on interaction dynamics and task-level failures.\\
\midrule
\cite{ahmad_2023} & No & Yes & Case Study & Collaboration patterns and role distribution & Our study also examines practical task execution and failure mechanisms during programming.\\
\midrule
Our Study & No & Yes & Observational (26) & Completion, failures, perception, quantitative analysis of abandonment & -\\
\bottomrule
\end{tabular}
\end{table*}

\subsubsection{Evaluating LLMs for Task Completion}

Existing research has examined task completion through both controlled experiments and in situ observations. 
    Controlled studies frequently use simple, benchmark tasks such as Leetcode-style problems or debugging exercises to measure completion rates, code quality, and efficiency \cite{meem2025software, chouchenHowWareDevelopers2024, imai2022github}. 
    Field studies, in contrast, capture how developers incorporate LLMs into complex tasks, revealing that while models accelerate progress on simple tasks, they often produce incomplete or overwhelming responses that require significant user intervention \cite{khojah2024beyond, sagdic2024taxonomy}. 
These findings highlight the variability that user behaviors introduce, along with differences in task demands, in task completion with LLMs. 
 We continue this line of work by contributing a detailed, empirical study to reflect real-world usage of LLMs, which lends way for a more nuanced understanding of user experiences.

Empirical studies in human-AI interactions in SE have found that LLMs can be helpful in speeding up the coding process by generating code. 
These works mainly focused on day-to-day work that software engineers are already familiar with. However, we know that software engineers commonly work on unfamiliar problems that require several steps to complete.
    Research with professional developers found that ChatGPT is often used to receive guidance, understand new topics, and clarify task requirements ~\cite{khojah2024beyond} found that ChatGPT can reduce communication in teams, while others observed that it provides both conceptual and technical explanations alongside basic code generation \cite{chouchenHowWareDevelopers2024}
    Practitioners similarly rely on ChatGPT for writing and debugging code \cite{meem2025software} and for interpreting ambiguous specifications \cite{sagdic2024taxonomy}.

We aim to complement the growing body of research seeking to understand how human behaviors shape the outputs and effectiveness of LLMs during  SE tasks.
While prior insights largely draw on cases where tasks were successfully completed with ChatGPT, we focus on open questions about when LLMs fail to support developers for task completion, especially when tasks are novel or ill-structured and unscaffolded.

\subsubsection{Evaluating LLMs for Helpfulness}
In our study, to understand the real-world usability of LLMs in SE practice, we examine task completion as an indication of helpfulness.
Many prior works' evaluation of AI-assisted programming emphasizes productivity and code correctness to suggest future development designs ~\cite{imai2022github, weisz2022better}. 
    Our work aims to highlight the importance of assessing how users complete tasks with LLM support, and their process of adapting from imperfect model responses. This further designates the need for task-oriented evaluations.
This reflects a broader HCI perspective that emphasizes user experience and perceived helpfulness in human–AI collaboration \cite{candon2022perceptions, lill2024helpfulness}.

Prior works that examine specific shortcomings of LLMs for SE focus on model failures, such as code quality~\cite{yetistiren2022assessing}, or task completion time~\cite{vaithilingam_2022}, as potential areas of focus for tool development aimed at reducing tool abandonment and improving user experience. 
Complementary insights from HCI show that users abandon intelligent systems when assistance becomes cognitively taxing, or untrustworthy \cite{candon2022perceptions, weisz2022better}.
Similar results have been observed in SE settings.
    Studies of open-source ecosystems find that contributors abandon pull requests due to task completion blocks, unclear feedback, or lengthy task cycles, resulting in lost opportunity costs, and often causing projects to become abandoned \cite{khatoonabadi2023wasted, khatoonabadi2023understanding}.
    With LLM tools, when faced with continuous, unhelpful answers, developers can frequently choose to abandon the tool, resulting in time and effort wasted \cite{sun2025don}. 
\textbf{Building on these findings, our study further explores LLM helpfulness by assessing the actual proportion of useful responses through observations and statistical testing. We also make the distinction of users choosing to abandon LLMs as opposed to the task, further contextualizing when user-LLM interactions result in wasted efforts. }
In our study, to better understand the failures and mitigations associated with LLMs for SE, we evaluate helpfulness as task completion and abandonment as a user perception to stop using tools, while making sure to note any frustrations that could compound user behavior and decision-making.

Beyond SE tasks, LLMs have demonstrated usefulness across various domains, lowering the barrier for acquiring new expertise. 
    Researchers highlight the importance of critical thinking when using LLMs~\cite{jeuring_2024, rasul2023role} and note varying levels of user trust and confidence~\cite{amoozadeh2024trust, ahmad_2023}. 
    For example, Levine et al.~\cite{levine2025students} investigated the role of ChatGPT in writing support, finding that its correct usage can enhance students' thought processes and writing quality. 
    Chen et al.~\cite{chen2024learning} examined differences in LLM-assisted learning between experts and novices in NetLogo, a programming language, and found that experts perceived greater benefits from LLM support than novices. 
    Choudhury et al.~\cite{choudhury2024impact} conducted an online survey and identified a correlation between lower perceived workload and increased user trust in ChatGPT, reinforcing the premise that user satisfaction predicts trust.

\subsection{Distinct Interaction Characteristics of LLM-based Tools}
Prior SE and HCI research on tool adoption and usability generally assumes that tools exhibit deterministic and stable behavior \cite{bruch2009learning, harman2012role}. Developers typically interact with fixed-function tools whose outputs remain consistent across uses, which allows existing models to focus on constructs such as perceived usefulness, ease of use, trust, and adoption intention \cite{bonhard2006accounting, elahi2021beyond, sinha2002role}. In contrast, LLM-based assistants generate multi-turn, adaptive, and stochastic outputs that evolve with each user prompt \cite{rasnayaka2024empirical, denny_2024, achiam2023gpt}. This generative behavior positions LLMs as co-participants in problem solving rather than static tools \cite{tian_2023, kazemitabaar_2024, barke_2023}, and it creates breakdown patterns such as cascading errors, context drift, and iterative misalignment that have no direct analogue in conventional SE tools \cite{xu_2022, prather_2024, heersmink2024use}.
While this work highlights important properties of LLM-supported development, it does not fully address how the interactive and evolving nature of LLM outputs differentiates them from traditional SE tools or how these characteristics influence failure and abandonment during task execution.

Because LLM failures unfold across turns rather than in isolated instances \cite{brown2020language, kuhail_2023, dang_2022}, abandonment in LLM-assisted workflows emerges as a dynamic behavioral response that develops during ongoing interaction. This differs from traditional technology adoption decisions, which are typically evaluated at a single time point or based on stable perceptions \cite{sinha2002role, Shani_Gunawardana_2010, bell2007modeling}. Existing survey-based studies and one-shot programming tasks provide limited insight into such real-time failure dynamics \cite{liang2024large, fan_2023}. These distinctions create a need for fine-grained empirical investigation into when failures occur, how users attempt to recover from them, and why some users ultimately disengage from the tool during task execution \cite{khurana2024and, schmidhuber_2021, prather_2024, li2025always}. These distinctions motivate our research questions, which examine how failures emerge, accumulate, and lead to user abandonment during multi-step SE tasks.

Prior work in both SE and HCI has documented many of the individual failure patterns we observed, such as ineffective prompting, loss of context, and mismatched user expectations \cite{khurana2024and, laban2025llms, rasnayaka2024empirical}. However, this body of work typically examines such issues in isolation, often focusing on correctness, usability, or productivity impacts. Our abandonment perspective instead treats these failures as interactional trajectories, showing how repeated breakdowns accumulate and ultimately lead users to disengage from LLM support altogether. This shift from isolated errors to abandonment as an outcome offers a complementary lens for understanding when and why LLM-based tools cease to be usable in practice.

\section{Methodology}
 
\subsection{User Study Design} \label{user_study}

\begin{figure}[t]
    \centering
    \includegraphics[width=0.7\columnwidth]{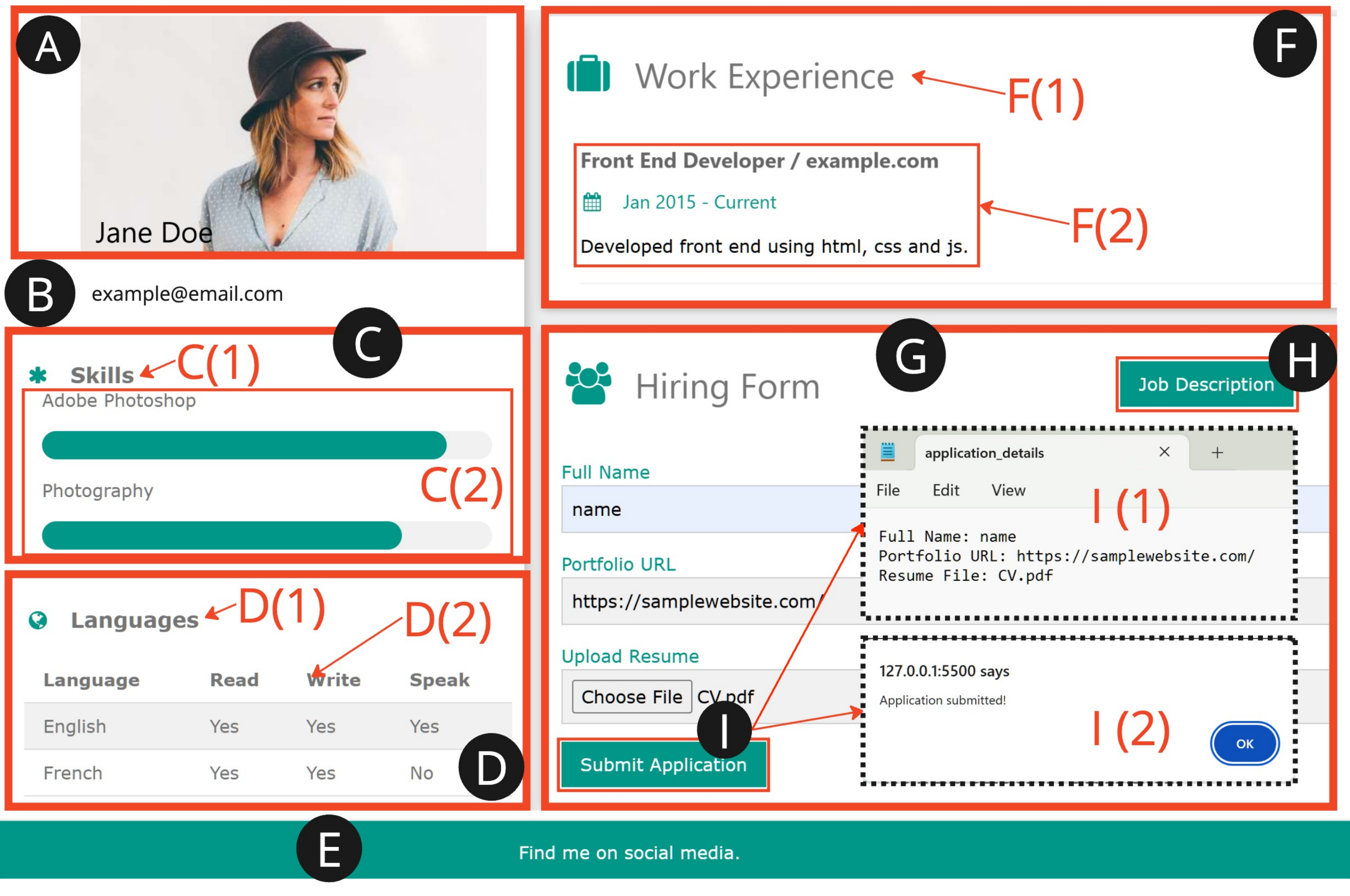}
    \vspace{-1em}
    \caption{Task Breakdown: \textcolor{red}{A:} Insert profile picture, \textcolor{red}{B:} Link email, \textcolor{red}{C(1):} Create division with headings/widgets, \textcolor{red}{C(2):} Add visualization, \textcolor{red}{D(1):} Create division with headings/widgets, \textcolor{red}{D(2):} Insert table, \textcolor{red}{E:} Insert footer division, \textcolor{red}{F(1):} Format side-by-side division, \textcolor{red}{F(2):} Insert headings/subtitles, \textcolor{red}{G:} Insert division with form/buttons, \textcolor{red}{H:} Implement pop-up on button click, \textcolor{red}{I(1):} Implement form with local file saves, \textcolor{red}{I(2):} Alert on submission.}
    \Description{Task breakdown diagram showing labeled steps A to I(2) as described in the caption.}
    \label{fig:interview_window}
\end{figure}
\subsubsection{Task Design} \label{task_design}
 
We designed a complex SE task to evaluate LLM-based assistants like ChatGPT, requiring multi-file edits, diverse languages, and visual integration. Structured as a web application project, the task was informed by real-world practices, senior-level SE courses, and industry consultations. It replicates a tutorial-based website to simulate a typical learning scenario.
To accommodate varying skill levels, the task progresses from HTML to CSS, JavaScript, and deployment, enabling participants to engage at an appropriate difficulty level while assessing LLM support across different development stages.  
Fig.~\ref{fig:interview_window} presents the expected outcome. 

Interviews with professional SDEs further confirmed that working with unfamiliar code, frameworks, or technologies is routine in industrial practice. This supports our choice to design the task around unfamiliar components, as such scenarios reflect common conditions under which developers rely on LLM-based assistance rather than representing novice-specific behavior.

Participants used ChatGPT-4 (accessed March-May 2023) during the study, chosen for its widespread adoption in programming~\cite{budhiraja2024s} and image-processing capabilities in front-end design~\cite{achiam2023gpt}. During the recruitment of SDEs to replicate the student-participant studies, GPT-4o (accessed May 2024) was released, leading the last five SDE participants to use GPT-4o ~\cite{gpto1} instead of GPT-4. While this transition introduces some variability, the study prioritizes representativeness in SOTA LLM techniques and aims to avoid reporting outdated issues. We discuss the potential limitations in Sec.~\ref{threats}.

While task complexity may amplify individual differences in prior experience, our analysis explicitly distinguishes user-related difficulties from LLM-related failures (RQ2), triangulates multiple data sources, and controls for coding experience in our quantitative models, thereby reducing the risk that observed interactional breakdowns are attributable primarily to participant proficiency.

\textbf{Evaluation rubric. } To objectively assess task completeness, two authors iteratively developed and  refined a structured rubric across three pilot studies, independently scoring each participant’s submission and resolving discrepancies through discussion until reaching consensus. The finalized rubric, consistently applied to all   participants, evaluates each subtask based on: (1) correct implementation of fundamental elements, (2) adherence to formatting, styling, and structural requirements, and (3) accuracy in functionality, alignment, and interactivity. 
This rubric ensures that both the content and design of the web pages meet the intended specifications. Using it, we quantified task completeness to enable consistent participant comparisons. 
 With 13 subtasks, each worth 3 points, the total possible score is 39. The complete rubric is available in the replication package~\cite{anonymous_2024_13179522}.
 
\begin{table}[t]
\centering
\footnotesize
\caption{Participant background and self-reported proficiency levels. Participants are grouped by study phase (students and professional SDEs) and by the LLM version used in the study (GPT-4, GPT-4o, and GPT-5.1).}
\label{tab:participant_skills}
\begin{tabular}{l>{\raggedright\arraybackslash}p{0.3\linewidth}llll}
\toprule
\textbf{ID} & \textbf{Primary Domain} & \begin{tabular}[c]{@{}l@{}}\textbf{Coding }\\\textbf{Exp (yrs)}\end{tabular} & \begin{tabular}[c]{@{}l@{}}\textbf{HTML }\\\textbf{skill}\end{tabular} & \begin{tabular}[c]{@{}l@{}}\textbf{CSS }\\\textbf{skill}\end{tabular} & \begin{tabular}[c]{@{}l@{}}\textbf{JS}\\\textbf{skill}\end{tabular} \\
\midrule

Pilot1 & Social Sciences & $<$ 3 & 2 & 1 & 1 \\

Pilot2 & Architecture & $<$ 3 & 1 & 1 & 1 \\

Pilot3 & SE & $>$ 5 & 4 & 3 & 3 \\

P4 & SE & 3-5 & 5 & 5 & 5 \\

P5 & SE & $<$ 3 & 2 & 2 & 3 \\

P6 & AI/ML research & $<$ 3 & 2 & 2 & 2 \\

P7 & SE, AI/ML research & $<$ 3 & 1 & 1 & 1 \\

P8 & AI/ML research & $<$ 3 & 1 & 1 & 1 \\

P9 & AI/ML research & $<$ 3 & 4 & 1 & 1 \\

P10 & SE & $<$ 3 & 2 & 2 & 2 \\

P11 & AI/ML research & 3-5 & 1 & 1 & 1 \\

P12 & AI/ML research & 3-5 & 4 & 4 & 4 \\

P13 & AI/ML research & $<$ 3 & 2 & 2 & 2 \\

P14 & SE & 3-5 & 3 & 3 & 3 \\

P15 & AI/ML research & 3-5 & 3 & 1 & 1 \\

P16 & AI/ML research & $<$ 3 & 2 & 2 & 3 \\

P17 & AI/ML research & $<$ 3 & 2 & 2 & 2 \\

P18 & AI/ML research & $<$ 3 & 1 & 1 & 1 \\

P19 & SE, Physics & $<$ 3 & 3 & 3 & 2 \\

P20 & Communication Engineering & $<$ 3 & 1 & 1 & 1 \\

P21 & SE & $<$ 3 & 2 & 2 & 2 \\

P22 & SE & $<$ 3 & 1 & 1 & 1 \\

P23 & Hardware & $<$ 3 & 1 & 1 & 1 \\

P24 & Electronic information engineering & 3-5 & 1 & 1 & 1 \\

P25 & SE, AI/ML research & $<$ 3 & 1 & 1 & 1 \\

\midrule[\heavyrulewidth]
\textbf{Professional SDEs using GPT-4o} \\
\midrule

P26 & SE & 3-5 & 5 & 4 & 4 \\

P27 & SE & $<$ 3 & 2 & 2 & 2 \\

P28 & SE & 3-5 & 3 & 3 & 3 \\

P29 & SE & $>$ 5 & 5 & 5 & 4 \\

P30 & SE & $>$ 5 & 4 & 4 & 4 \\

\midrule[\heavyrulewidth]
\textbf{Professional SDEs using GPT-5.1} \\
\midrule
P31 & SE & $>$ 5 & 4 & 4 & 4 \\

P32 & SE & $>$ 5 & 4 & 4 & 4 \\

P33 & SE & $>$ 5 & 5 & 5 & 5 \\

P34 & SE & $>$ 5 & 4 & 3 & 4 \\

P35 & SE & $>$ 5 & 4 & 4 & 4 \\

P36 & SE & $>$ 5 & 3 & 3 & 4 \\
\bottomrule

 \end{tabular}
\end{table}

\subsubsection{Recruiting Participants}\label{recruitment}

The study initially used a lab setting with student participants, commonly recruited for SE experiments~\cite{soloway2009empirical, tahaei2022lessons, tahaei2022recruiting}. Prior research suggests post-graduate students closely match professional SDEs with marginal differences, making them a prerequisite for studies before SDEs. ~\cite{tichy2000hints}.
To explore potential transferability and strengthen result validity, we extended recruitment to professional software developers (SDEs), enabling a comparative analysis across experience levels. 
As LLM capabilities continue to evolve and newer models supersede earlier versions, we additionally recruited six professional SDE participants to replicate the same study protocol using GPT-5.1~\cite{OpenAI_2025}.
Tab.~\ref{tab:participant_skills} details the participant background and self-reported experience levels.

Student participants were post-secondary students with programming experience with minimal web development exposure, simulating real-world scenarios where SDEs navigate tasks beyond their expertise.\footnote{To ensure the scenario accurately reflected real-world workflows, we validated with SDEs that it aligned with tasks requiring an initial learning phase before execution.} For SDE participants, we recruited those with professional development experience. Participants were recruited via university mailing lists and alumni networks, supplemented by convenience sampling, where participants were encouraged to refer peers who met our selection criteria. 
All participants signed up via a Google Form, providing background information, web development, and LLM experience.

\begin{table}[t]
\centering
\footnotesize
\caption{Summary of Participant Experience with LLMs; in total 36}
\label{tab:participant_llm}
\begin{tabular}{llll}
\toprule
\textbf{Familiarity with LLMs} & \textbf{\#Students w/ GPT-4} & \textbf{\#SDEs w/ GPT-4o} & \textbf{\#SDEs w/ GPT-5.1}\\
\midrule
I do not know what LLMs are & 4 & 0 & 0\\
I have heard of LLMs & 5 & 1 & 0\\
I vaguely know how LLMs work & 11 & 1 & 2\\
I can clearly explain what an LLM is & 5 & 3 & 4\\ 
\midrule[\heavyrulewidth]
\textbf{Experience with LLMs} & \textbf{\#Students} & \textbf{\#SDEs w/ GPT-4o} & \textbf{SDEs w/ GPT-5.1} \\
\midrule
I have never used LLMs & 10 & 1 & 0\\
I have used LLMs before & 15 & 4 & 6\\
\bottomrule
 \end{tabular}
\end{table}

\textbf{Participant Overview and Experience.}
We recruited 30 participants, including three for pilot studies. 
Following standard empirical software engineering practice, three pilot sessions were conducted solely to refine the study protocol and were excluded from the final analysis.
Of the remaining 27 participants, we excluded P10 due to significant deviations from the task instructions, yielding a final dataset of 26 complete study units.
Tab.~\ref{tab:participant_llm} presents the participant's prior knowledge and experience with LLMs. 
Most student participants had less than three years of coding experience, limited proficiency in HTML, CSS, and JavaScript, and only a basic familiarity with LLMs from brief usage.  
SDEs had between three and over five years of experience, ranging from familiarity to proficiency in HTML, CSS, and JavaScript. Most had prior exposure to LLMs and were familiar with their use.  
While participants varied in professional experience, our analysis focused on recurring interaction-level breakdown patterns rather than individual proficiency. These breakdown categories were observed across both student participants and practicing SDEs, indicating that the identified pitfalls are not limited to novice-level domain knowledge.
We validated with SDEs whether the scenario accurately reflects their daily workflow—specifically, tasks that require an initial learning phase before execution.

\subsubsection{Study Protocol}
All studies were conducted virtually via Zoom, where participants controlled the experimental screen. With consent, we recorded audio and screen activity. The split-screen setup included Google Chrome (showing the target webpage \& ChatGPT) and VSCode (codebase with real-time updates).  
Participants followed a think-aloud protocol~\cite{ku2010metacognitive}, enhancing insight into their reasoning while promoting metacognition~\cite{garner1989metacognition}. 

Before taking control, participants received a task overview and brief ChatGPT introduction. They could search online, take screenshots, and upload files as part of their queries. After the briefing, they navigated the task autonomously, simulating real-world independent exploration.  

To capture realistic tendency to abandon ChatGPT, participants had unlimited time. Pilot studies showed they completed or stopped within ~1 hour (avg. 45 min). The session concluded when participants either achieved their goal or discontinued the task, ensuring that outcomes reflected human-LLM interactions rather than individual differences.  

\textbf{Post-Task Reflection.} After each session, participants shared feedback via a semi-structured interview.
Three participants (P4, P5, P6) answered the interview questions through text format due to timing. The following questions guided their responses, offering detailed insights into their experiences with ChatGPT during the study session. This method captured user perspectives that might not have been fully expressed during programming, as well as their expectations for future interactions with ChatGPT.
 The study protocols~\cite{anonymous_2024_13179522} were approved by the research ethics board at the institutions of the authors who participated in conducting the interview studies.

Note that while we provided an overview of how to use ChatGPT, we intentionally did not provide participants with formal prompt-engineering training prior to the task. Our goal was not to assess LLM performance under expert-optimized prompting, but to examine how failures arise during realistic, low-scaffolded tasks in human–LLM interaction for SE tasks. Prior work shows that developers, including professionals, often lack stable prompt engineering skills and rely on iterative trial-and-error rather than optimized prompting strategies~\cite{kruse2024can}. Providing training would have altered interaction dynamics and reduced visibility into early-stage failure that practitioners frequently encounter during real-world adoption. Accordingly, prompt-related breakdowns are treated as ecologically valid user-centric interaction failures rather than experimental artifacts.

\begin{framed}

\noindent \textbf{Reflection Questions}
\begin{enumerate}[leftmargin=*,label=Q\arabic*., itemsep=0.5ex, parsep=0pt]
    \item Were there any points where you wanted to give up on using ChatGPT for completing the task? Why? 
    \item What was the main problem with ChatGPT’s answers?
    \item If you were to continue this task, would you use any other resources?
    \item Rate ChatGPT’s responses from 1 (not at all helpful) to 5 (very helpful). Why?
    \item Rate your understanding of this code from 1 (no understanding) to 5 (very familiar). Why?
    \item Do you feel more confident about using HTML, CSS, and JavaScript in the future?
    \item (If applicable) What did you feel about ChatGPT breaking down the steps of the answers for you?
    \item Was there anything I didn’t cover that you would like to share?
\end{enumerate}
\end{framed}

\subsection{Analyzing Multi-Source Qualitative Data (RQ1-4)} 
\label{qual-analysis}

\subsubsection{Pilot Development and Reliability}
Each study session was analyzed immediately after completion. 
The first three sessions served as pilot studies to iteratively develop and refine a structured rubric.
Inter-coder reliability improved across rounds: Cohen’s $\kappa$ increased from 0.78 (Round~1) to 0.86 (Round~2), reaching 1.0 in the final pilot once code definitions were clarified. 
     This level of agreement reflects the stabilization of the coding scheme rather than independent blind agreement, which is consistent with qualitative pilot refinement practices. 
    After the pilots, the finalized codebook was used for all independent coding followed by consensus discussions.

\subsubsection{Coding Procedure and Thematic Analysis}

\paragraph{\textbf{Defining QnA Units.}}
For the interaction data, we decomposed ChatGPT usage into atomic interaction units, where each unit consists of a participant prompt or question and the corresponding ChatGPT response. We refer to each such prompt–response pair as a \textbf{QnA} unit. These units served as the primary analytic unit for coding interaction-level failures, mitigation strategies, and outcomes, and were analyzed in conjunction with participants’ verbalizations and behavioral traces.

\paragraph{\textbf{Data Triangulation Across Modalities.}}
Two authors independently coded the data, comprising transcripts from think-aloud procedures and post-study reflections, as well as interaction logs and user action recordings. 
They employed data triangulation~\cite{jick1979mixing} by comparing observations across these multiple data sources. 
Examples of triangulated abandonment events, including timestamps, chat-turn counts, and think-aloud indicators, are presented in Sec.~\ref{mitigation}.
Cross-source comparison helped confirm cases where participants perceived a response as useful yet still abandoned it during implementation, and prevented over-reliance on self-report alone.

\paragraph{\textbf{Inductive Thematic Analysis Procedures.}}
Using inductive thematic analysis (following Braun \& Clarke's guideline~\cite{Braun_Clarke_2024}) and adapted to the multimodal interaction data. 
Our analysis integrated think-aloud transcripts, post-task reflections, screen recordings, and interaction logs to capture both participants’ reasoning and observable interaction outcomes. Initial codes were generated through line-by-line inductive coding of transcripts and interaction traces, using researcher-defined analytic labels to summarize observed behaviors and breakdowns. 

Two labellers independently coded an initial subset of sessions, iteratively refined the shared codebook, and resolved discrepancies through discussion until consensus was reached. As coding progressed, related codes were clustered into higher-level categories and subsequently synthesized into themes that captured recurrent interactional mechanisms aligned with our RQs.

\paragraph{\textbf{Cross-Source Coding Examples.}}
To illustrate how these procedures were applied in practice, we provide several examples of how verbal statements and behavioral traces were integrated during coding.
Examples of coding included: (1) recoding ambiguous think-aloud utterances by checking the corresponding on-screen action; (2) resolving mismatches between a participant’s stated plan and the logged sequence of edits; and (3) clarifying whether a ChatGPT output was ``helpful'', ``partially helpful'', or ``unhelpful'' by reviewing recorded footage of participants' code acceptance and follow-up action. If participants left the code in the codebase, this was considered ``helpful'', if the participant left parts of the code in, or engaged in mitigating behavior, this was considered ``partially helpful'', and if participants removed all code, or restarted, this was considered ``unhelpful''. If participants later replaced the code with an updated version, this did not affect the helpfulness rating in the moment, as participants in the moment found the code to be helpful. These procedures ensured consistent interpretation across sources.

\paragraph{\textbf{Classifying Failure Sources.}}
Building on this coding framework, we then conducted a focused analysis to classify the contributing mechanisms of task completion failures.
Once interaction failures were identified (RQ1), we analyzed their underlying causes to distinguish those attributable to users from those stemming from ChatGPT. 
Many failures resulted from multiple factors. For example, insufficient detail in a user’s prompt (user-related) could combine with ChatGPT’s limited contextual understanding (ChatGPT-related). 
We approached differences between the user- and ChatGPT-related mechanisms by systematically analyzing failure cases, identifying recurring patterns, and categorizing them based on whether each issue originated from user input, ChatGPT’s response, or a combination of both.

To increase transparency, we applied explicit decision criteria during coding. A failure was labeled user-contributed when missing constraints, ambiguous instructions, or deviations in user actions directly preceded the breakdown. A failure was labeled ChatGPT-contributed when the prompt sufficiently specified task requirements, but the generated output was incorrect, incomplete, or inconsistent with those requirements. When both factors were evident, we coded the case as interactional. These categories function as analytical distinctions for characterizing contributing mechanisms rather than experimentally isolated primary causes.

To identify the interaction mechanisms that contributed to the failure, we compared across different chatlogs in ChatGPT in different contexts for similar prompts and tasks. For similar prompts, we expected that ChatGPT should give similar quality of answers, and made note when it changed, such as missing preconditions, or mismatches between the expected output from natural language explanations by ChatGPT and actual output from generated code. If the prompt structure and goal were similar (ie. ``how do I do this task?'' compared with ``what should the code look like for completing this task?'') and the output differed, such as one failing, while the other did not, we compared across the chat history to understand the key mechanisms for the failure. Additionally, we also manually read through each answer to familiarise ourselves with the helpfulness of the generated results, making note of potentially helpful/unhelpful segments. We combine this with observations from user actions. If user actions deviated from other actions that resulted in failures, we logged those as user mechanisms for failure.
This classification allowed us to examine the interplay between user behaviors and system limitations, providing a comprehensive perspective on the factors shaping interaction outcomes.
 
\paragraph{\textbf{Identifying Abandonment Events.}}
 To determine when participants abandoned ChatGPT during the task, we analyzed multiple forms of evidence. We first examined interaction traces and participants’ think-aloud notes to identify discontinued prompting, extended pauses, or explicit statements indicating a decision to stop using the tool. We then corroborated these behavioral indicators with participants’ post-task reflections. To further improve the rigor of this measurement, we incorporated additional interaction data, including prompt histories, timestamps, and prompt frequency patterns, which helped distinguish transient frustration from genuine disengagement. 

We used temporal markers and verbal cues to mark momentary frustration, such as short pauses before renewed prompting, thinking about prompt formulation through paused action, and longer pauses with think-aloud processing. These pauses were combined with any user behaviors, such as long sighing, verbal cues of frustration (such as users complaining about code output), to form momentary confusion. Notably, genuine disengagement may be preceded by these actions. To further distinguish genuine disengagement, we observed prolonged inactivity, including lack of think-aloud processes

 This was especially prevalent when users disengaged with LLM for multiple tasks, regardless of time spent away. We triangulated observation points of users' think-aloud processing that signaled disengagement (such as saying that they do not want to use ChatGPT anymore), users' actions (multiple tasks completed without ChatGPT interactions
 and prompt patterns (reduced interactions, worsening perceptions of the quality of answers revealed in think-aloud and follow-up prompts).  

 This multi-source validation allowed us to more reliably identify abandonment events and provided a stronger foundation for our quantitative analysis in RQ5.

\paragraph{\textbf{Saturation and Scope of the Dataset.}}
Given the heterogeneity of our sample (students and SDEs) and the use of multiple model versions, we explicitly monitored thematic development throughout data collection.
After analyzing data from 17 student participants, we found that the additional four students and five SDEs contributed only marginal insights. While they deepened our understanding of existing themes, no new themes emerged. 
This confirmed that our analysis provided sufficient depth to address our RQs within the studied contexts, and that the additional SDE sessions served primarily to corroborate existing interactional patterns rather than introduce new ones~\cite{guest2006many,francis2010adequate}. 
We did not observe novel behaviors from the SDEs that set them apart from the student participants. 
We therefore interpret our findings as having reached code-level thematic saturation within the scope of the studied contexts. At the same time, we acknowledge that saturation is bounded by the characteristics of our sample and task setting, and future work with different developer populations or interaction contexts may surface additional themes.

\paragraph{\textbf{Methodological Rationale. }} To justify our analytical choices, we selected this structured approach to better align with the nature of our data.
Because our study involved multimodal and fine-grained interaction data, including think-aloud transcripts, screen recordings, and prompt logs, we adopted an iterative, trace-informed qualitative coding approach rather than a purely narrative or reflexive thematic analysis. This approach allowed us to systematically align observable user behaviors with verbalized reasoning, while preserving transparency and reproducibility in the coding process. Structured coding was particularly suited to capturing temporal and behavioral phenomena such as error recovery, mitigation, and abandonment, which are difficult to represent using purely interpretive methods. As is standard in mixed-methods human–AI interaction research, our claims focus on the transferability of observed mechanisms rather than population-level prevalence.

\paragraph{\textbf{GPT-5.1 Replication Analysis.}} We analyzed the GPT-5.1 sessions using the same finalized codebook, triangulation strategy, and analytic procedures described above. The purpose of this replication was to examine the robustness of the identified failure categories, mitigation strategies, and abandonment dynamics in a newer model rather than to generate additional themes or compare model performance. 
To ensure transparency in our saturation claims across heterogeneous samples and model versions, we applied the same saturation monitoring procedure used in the primary dataset. Specifically, we tracked the emergence of new codes per interview using a saturation log and operationalized code saturation as the absence of new codes across three consecutive interviews.
Coding of the GPT-5.1 sessions yielded the same failure taxonomy and interactional patterns, and no new themes emerged, indicating that the analytic framework transferred consistently across model versions. 
We did not observe qualitatively distinct behaviors attributable to model version differences.
To strengthen analytic trustworthiness, we conducted member checking after coding by discussing interpretations and conclusions with participants and resolving discrepancies. 
Consistent with established qualitative saturation criteria~\cite{braun2021saturate}, interviews after the third GPT-5.1 participant yielded only marginal refinements and no new themes; we therefore concluded data collection after six interviews.

\subsection{Hypothesis Testing (RQ5)} \label{hypo} 
We identified context factors influencing ChatGPT abandonment, formulated four hypotheses based on prior research, and operationalized measures. Using data from 26 study sessions, we conducted a logistic regression analysis to test these hypotheses.
The regression was estimated using a binomial family (logistic link) and included fixed effects for both participant ID (id) and task (task), and were automatically clustered at all fixed-effect variables ~\cite{abadie2023should}.

\subsubsection{Identifying Potential Context Factors}
Understanding the factors that influence a user's likelihood of disengagement when interacting with AI-assisted programming tools is critical for improving user experience and productivity. Based on prior research in SE, human-computer interaction, and problem-solving behavior ~\cite{etcuban2025use, shah2023continuance, bosch2017affective, chiou2023trusting, bandura1977self, soloway2009empirical}, we propose the following hypotheses:

\textbf{H$_1$: The helpfulness of ChatGPT's responses is negatively correlated with the likelihood of giving up. }
    When ChatGPT provides helpful responses, users gain clearer guidance, reducing cognitive load and increasing confidence in solving the problem ~\cite{etcuban2025use}. 
    Prior research on AI-powered programming support has shown that useful feedback sustains user motivation and fosters persistence in problem-solving tasks ~\cite{shah2023continuance}. 
    Conversely, when responses are unhelpful, users may experience frustration and uncertainty, making them more prone to disengagement.

    \noindent

\textbf{H$_2$:  Higher task performance scores are associated with a lower likelihood of giving up. }
    Task performance serves as an indicator of a user’s success in completing a given task, aligning with self-efficacy theory, which posits that individuals with greater perceived ability are less likely to disengage ~\cite{bandura1977self}. 
   Prior work demonstrated that early success in problem-solving is associated with sustained effort and perseverance, while lower performance may lead to frustration and eventual task abandonment ~\cite{zeller2009programs}.

    \noindent

\textbf{H$_3$: A greater number of prompts used is negatively correlated with giving up.}
    A higher number of prompts suggests continued engagement in the problem-solving process.
    Iterative refinement is an established strategy in debugging and programming, where persistence often leads to improved solutions ~\cite{zeller2009programs}. 
    Research on interactive AI agents suggests that responsive agents and users exhibit higher resilience, whereas those who disengage early tend to abandon tasks more frequently ~\cite{chiou2023trusting}. 
    This pattern suggests that prompt iteration could be an indicator of persistence rather than frustration-induced disengagement.

\textbf{H$_4$: A higher coding experience is negatively correlated with giving up.}
   Experienced programmers possess stronger problem-solving skills, making them more resilient when encountering challenges ~\cite{soloway2009empirical}. 
   Studies in computer programming education suggest that novice developers are more susceptible to frustration and disengagement when faced with difficulties, while more experienced programmers are better equipped to navigate obstacles and continue working toward a solution ~\cite{bosch2017affective}.

\looseness=-1

\subsubsection{Operationalizations.} \label{factors}
We iteratively designed the measures for contextual factors and control variables. 

\textbf{Outcome: Tendency to Abandon ChatGPT for Task Completion.}  
After the task, participants answered: ``At which points did you want to give up on using ChatGPT? Why?'' (Q1). If they considered abandoning ChatGPT, we coded it as \(1\); otherwise, \(0\).  
We assessed abandonment at least once without linking it to specific subtasks, as real-time questioning would have been impractical. Instead, post-session reflections provided a feasible alternative, though this coarse granularity may impact results (discussed in Sec. \ref{threats}).  

 Although the post-task interview question served as the binary label for the regression model, it was not used in isolation; each instance was cross-validated using interaction timestamps, prompt cessation, and think-aloud indicators to ensure it reflected genuine disengagement rather than momentary frustration.

\textbf{Measurement for Helpfulness.} We objectively assessed helpfulness by analyzing recorded interactions and screen activities during each task session. Specifically, we classified ChatGPT responses based on whether the participants successfully implemented the generated code or utilized the provided explanations to progress their tasks.

\begin{itemize}[leftmargin=1.5em]
    \item \textbf{Helpful}:  Fully addressed the query, enabling direct task completion without further modifications.
    \item \textbf{Partially Helpful}: Provided relevant information but required additional effort, such as prompt refinement or scaffolding.
    \item \textbf{Unhelpful}: Failed to assist task progress, due to misleading, incomplete, or irrelevant information.
\end{itemize}

\noindent
\textbf{Measurement for  Product Completeness.} 
As outlined in  Sec.~\ref{task_design}, we evaluate participants' web pages for each identified subtask (shown in Fig.~\ref{fig:interview_window}) using a structured rubric.  Participants sometimes ask ChatGPT a question that spans multiple subtasks; in such cases, we assess each subtask separately using the rubric and sum the scores for the session.

\textbf{Control Variables.} 
We controlled for two variables that may influence the outcome: (1)  participant's background (student or SDE) and (2)  years of coding experience. These were selected because participants' expertise and familiarity with programming could impact their interaction behaviors and task outcomes, potentially confounding our analysis of abandonment tendencies.

\begin{figure}[t]
    \centering
    \includegraphics[width=0.7\columnwidth]{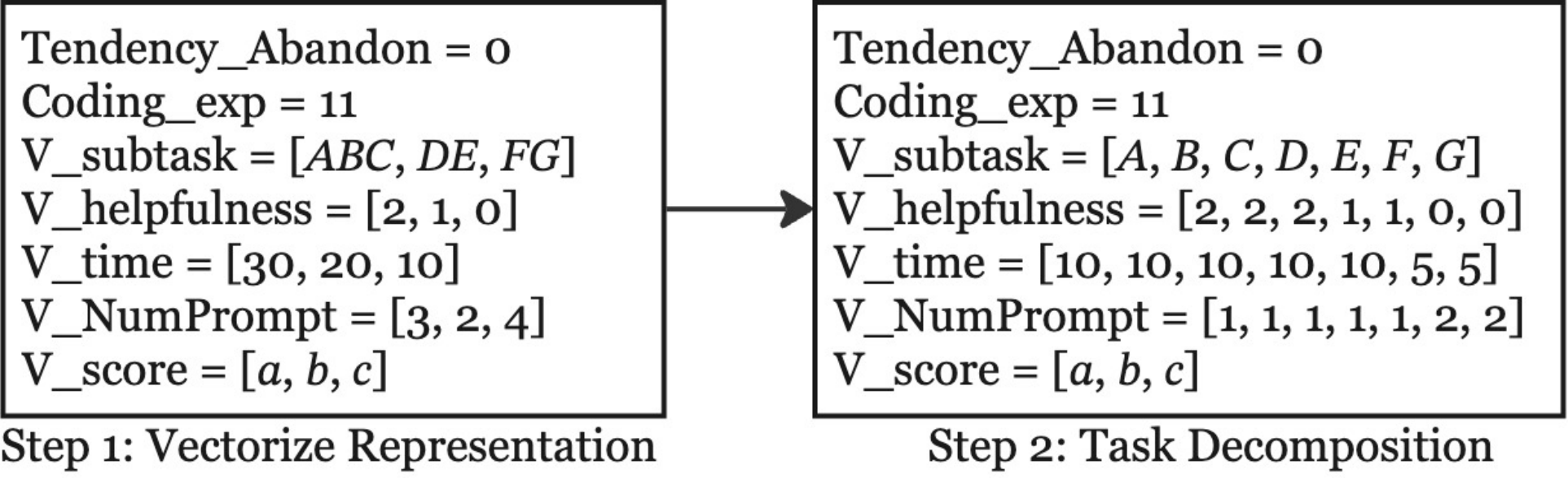}
    \caption{Illustration of vectorizing P26's session data.}
    \Description{The vectorization process, with step one creating a vectorized representation of a participant's data, and step two being the task decomposition of the vectorized data.}
    \label{fig:vector_fig}
\end{figure}

\textbf{Vectorization.}
For each participant, we first segmented their coding process by subtask and examined the associated contextual factors. We then constructed a vector representation mapping each participant’s sequential interactions with ChatGPT into a multidimensional space, capturing subtask decomposition, task completion, and corresponding contextual factors.
For example, P26, an SDE with 11 years of coding experience, structured their interactions with ChatGPT into three subtasks (ST): ABC, DE, and FG. ST1 was partially helpful, involving two prompts over 30 minutes; ST2 was unhelpful, with two prompts over 20 minutes; and ST3 was helpful, requiring four prompts over 10 minutes. During the post-study discussion, P26 indicated they did not want to abandon ChatGPT during task completion.
Consequently, the vector representations for P26 are shown in Fig.~\ref{fig:vector_fig} (Step 1).

Since our study's unit of analysis is the individual task, when participants asked a question involving multiple subtasks, such as ``ABC,'' we disaggregated the response into its constituent tasks ($A, B, C$). Given that scores and time were reported as aggregate values, we distributed these metrics equally across the subtasks using an average-based approximation. Fig.~\ref{fig:vector_fig} (Step 2) illustrates the updated vectorization result for P26. This approach enables a granular, task-level dataset for subsequent analyses.

\subsection{Statistical Analysis}
To test these hypotheses, we employed logistic regression to 
test the outcome, whether it is significantly associated with the different hypothesized context factors, while controlling for confounding variables.
 After completing the first-order logistic regression estimates, we then further investigate the interactions to determine whether significant effects differ between students and SDEs.
We assessed multicollinearity using the variance inflation factor (VIF) and applied corrections as needed. 
For each model variable, we report the exponentiated coefficient (odds ratio), standard error, and significance level (p-value). The odds ratio represents the factor by which a one-unit increase in a predictor affects the odds of the outcome occurring—an increase if greater than 1 and a decrease if less than 1.

\subsection{Threats to Validity and Limitations} \label{threats}
\emph{Internal validity} may be affected by variations in participants’ experience with web development and LLMs, influencing response evaluation.  We intentionally did not provide formal prompt-engineering training in order to observe naturalistic usage patterns; however, this choice introduces a potential internal validity consideration. Some breakdowns attributed to ``LLM failures'' may partially reflect differences in participants' prompting practices rather than intrinsic model limitations. While we analytically distinguish user-contributed (UC) and ChatGPT-contributed (CC) factors, these categories are interpretive rather than experimentally isolated. Accordingly, our findings should be interpreted as characterizing human–LLM interaction failures under minimally guided usage conditions rather than isolating model-only limitations. Controlled studies manipulating prompting expertise would be needed to make stronger causal attributions.

\emph{External validity} is a limitation as we only studied ChatGPT. 
Our work exhibits the typical threats common and expected for this kind of qualitative, observational research. Generalizations beyond the sampled participant distribution should be made with care.
This concern is amplified by our limited sample of professional developers; while we included a small sample of professional developers, larger-scale replication is needed to assess the extent to which these interactional patterns transfer to broader industrial contexts. 
Moreover, as is standard in mixed-methods human–AI interaction research, our claims focus on the transferability of observed interactional mechanisms rather than population-level prevalence or expert upper-bound performance.

\emph{Construct validity } concerns arise from the coarse granularity of measuring the tendency to abandon ChatGPT. 
Additionally, our vectorization process into measurable components and rubric-based assessment may not fully capture qualitative aspects.  
To address these concerns, we iteratively refined the rubric through pilot studies and ensured consistency in evaluation. Our measures capture only specific facets of quality, and results should be interpreted within these operationalized contexts.  

Our operationalization of helpfulness reflects whether a model response advanced task progress within the interaction. In practice, perceived progress may depend not only on the technical correctness of the model output, but also on participants’ ability to adapt that output into their workflow, as well as on contextual factors of the task.
Accordingly, helpfulness should not be interpreted as an isolated indicator of model performance, but rather as an interaction-level construct situated within a specific developer–model–task configuration.

In addition to these measurement considerations, we sought to improve transparency and reproducibility; we provide an R notebook detailing our analysis~\cite{anonymous_2024_13179522}. 
Lastly, since SDEs used GPT-4o while student participants used GPT-4, differences in outcomes could be influenced by model variation. However, given the strong consistency observed across responses, we cannot fully rule out this as a confounding factor. This reinforces the need for cautious interpretation of model-specific effects. 
Notably, in a follow-up study using GPT-5.1, we observed behavioral patterns similar to those in the original study. While this does not eliminate the possibility of model-related confounds, it provides additional evidence that the interaction patterns we report are not tied to a single model version.

\section{RQ1: Failures during ChatGPT Interaction }\label{sec:failure}

We define a failure case as a scenario in which ChatGPT's response fails to effectively support task completion, based on the previously introduced definition of helpfulness (Sec.~\ref{factors}). Specifically, a response is considered a failure if it is classified as either \emph{partially helpful} or \emph{unhelpful} for supporting task completion, as these cases require additional effort from users or provide no meaningful assistance. 
As such, task completion failures are not directly attributable to either ChatGPT or the users' own lack of knowledge, but rather an interaction breakdown. Thus, failures are a manifestation of the joint contributing mechanisms of users' prompt and integration skills and model output quality.

Tab.~\ref{tab:failures-doublecolumn} summarizes the failure cases we identified. 
These failures fall into three broad categories: (1) incomplete or incorrect responses, (2) cognitive overload due to information mismanagement, and (3) lack of context retention leading to redundant effort. Below, we explore how these issues manifested and their impact on user experience.

While some failures were influenced by prompt quality, a substantial portion stemmed from limitations in the model’s reasoning, context retention, and code synthesis. These patterns persisted even when prompts were well-formed, indicating that the observed failures cannot be fully explained by poor prompt engineering alone. Importantly, we disentangle failures attributable to user input quality from those due to inherent LLM limitations to avoid conflating prompt engineering issues with genuine model constraints.

\begin{table*}[t]
\centering 
\footnotesize
 \caption{Comparison of LLM failure categories across model versions. Counts and percentages show how the prevalence of specific failure types differs between legacy models (GPT-4 / GPT-4o) and GPT-5.1, while the overall failure taxonomy remains consistent.}
\begin{tabular}{@{}p{1.8cm}p{2.2cm}p{5cm}p{1.3cm}p{1.3cm}@{}}
\toprule
\makecell[tl]{\textbf{Category}} & 
\makecell[tl]{\textbf{[Index] Failure}} & 
\makecell[tl]{\textbf{Description}} & 
\makecell[tl]{\textbf{GPT-4 (n, \%)} }&
\makecell[tl]{\textbf{GPT-5.1 (n, \%)}}\\[-1ex]
\midrule

\multirow{4}{=}{\makecell[l]
{\textbf{Incomplete or }\\\textbf{Incorrect}\\\textbf{Responses}}}  
& \textbf{[F1]} Incomplete answer & ChatGPT omitted details when processing entire tasks in a single prompt (e.g., screenshots or codebases), leading to missing components. & 36 (40\%) & 10 (25\%) \\\cmidrule{2-5}

& \textbf{[F3]} Missing preconditions& ChatGPT failed to specify necessary setup steps before implementation, leading to incomplete solutions. & 19 (22\%) & 6 (15\%)\\\cmidrule{2-5}
& \textbf{[F7]} Erroneous answer & ChatGPT introduced errors or bugs into the generated code or responses. & 5 (6\%) & 2 (5\%)\\
\midrule

\multirow{4}{=}{\makecell[l]{\textbf{Cognitive }\\\textbf{ Overload}\\\textbf{\& Information}\\\textbf{Mismgmt.}}}  
& \textbf{[F2]} Overwhelming answer & 
Responses to multi-task queries were excessively long and lacked clear step-by-step breakdowns.
& 21 (24\%) & 5 (12.5\%) \\\cmidrule{2-5}
& \textbf{[F4]} Overcomplicated answer & Responses contained unnecessarily complex solutions when simpler approaches were available. & 19 (22\%) & 1 (2.5\%) \\\cmidrule{2-5}
& \textbf{[F8]} Incorrect image generation & Requested visual outputs were inaccurate or failed to reflect the intended content. & 3 (3\%) & 2 (5\%)\\
\midrule

\multirow{4}{=}{\makecell[l]{\textbf{Lack of Context }
\\\textbf{Retention}\\\textbf{\& Redundant }\\\textbf{ Effort }}}  
& \textbf{[F5]} Lacks context & ChatGPT failed to maintain continuity in discussions, leading to overly general or disconnected responses. & 24 (17\%) & 7 (17.5\%) \\\cmidrule{2-5}
& \textbf{[F6]} Irrelevant answer & ChatGPT misunderstood queries, producing off-topic or misleading responses. & 7 (8\%) & 10 (25\%)\\\cmidrule{2-5}
& \textbf{[F9]} Unresponsive to prompt changes & Despite user modifications, ChatGPT continued generating the same responses. & 16 (9\%) & 15 (37.5\%)\\
\bottomrule
\end{tabular} 
\label{tab:failures-doublecolumn}
\end{table*}

\subsection{Incomplete or Incorrect Responses}
One major challenge was \textbf{incomplete answers (F1, 40\%)}, where interactions with ChatGPT resulted in answers that omitted crucial details when handling large prompts. This issue was common when users provided entire codebases or screenshots, expecting a comprehensive response. Instead, ChatGPT produced partial solutions, requiring users to manually fill in missing components. 
For example, P9 uploaded a full code file but had to compare ChatGPT’s response to the expected website components, as the AI failed to generate certain sections.

Another frequent issue was 
\textbf{missing preconditions (F3, 22\%)}, where ChatGPT answers failed to specify essential setup steps before implementation. This was particularly problematic for users unfamiliar with debugging. For instance, P8 encountered a deployment error because ChatGPT omitted a required package installation. Since P8 lacked troubleshooting experience, they had to engage in multiple re-prompts to identify the missing step.

Similarly, \textbf{erroneous responses (F7, 6\%)} introduced further inefficiencies, as ChatGPT generated incorrect or buggy code. P25 requested code to insert a footer division, but ChatGPT mistakenly positioned it as a column instead. As P25 noted, ``\emph{I think manual editing would be better than prompting ChatGPT,}'' highlighting how AI-generated errors can increase workload rather than reduce it.

\subsection{Cognitive Overload \& Information Mismanagement}
While some users struggled with missing information, others faced
\textbf{overwhelming responses (F2, 24\%)}, where ChatGPT provided excessively long answers lacking a structured step-by-step breakdown. 
This was an interaction failure, particularly salient during debugging, where users ask for edits on a specific portion of the code, but received answers that regenerated the entire codebase.
This led to \textbf{cognitive overload}, forcing users to sift through dense outputs to extract relevant content. P4 described this frustration, stating, ``\emph{I wanted to give up when I had to parse through the code to find where I needed to make those changes.}''
Some users copied entire responses verbatim, hoping they were correct, but this approach rarely succeeded, leading to further debugging.

A related issue was \textbf{overcomplicated answers (F4, 22\%)}, where ChatGPT suggested unnecessarily complex implementations instead of straightforward solutions. For example, P17 asked how to change the text size, and ChatGPT recommended using the CSS \texttt{nth-child selector}, an advanced technique that was unnecessary for the task. A simple inline CSS rule would have sufficed. However, ChatGPT did not justify its choice, leaving P17 confused about whether the suggested approach was optimal.

Additionally, \textbf{incorrect image generation (F8, 3\%)} affected users working on front-end tasks. When P4 and P11 requested visual representations of their code, ChatGPT generated misleading or incomplete images rather than acknowledging its inability. 

\subsection{Context Loss \& Redundant Effort}
Several failures arose from ChatGPT’s limited memory, requiring users to repeat or fill in missing details.

For instance,  \textbf{responses lacked context (F5, 17\%)} led to workflow disruptions. P14 initially asked for HTML and CSS for inserting a profile picture. Later, when they requested specific values for ``\emph{width, top, left, and min-height}'', ChatGPT failed to reference its previous response, assuming a new context. This required P14 to reintroduce prior details manually.

Similarly, \textbf{irrelevant (F6, 8\%)} resulted from ChatGPT misinterpreting ambiguous queries. P6 and P21, for example, used the word ``\emph{block}'' to refer to webpage divisions, but ChatGPT misunderstood and generated unrelated solutions. Since ChatGPT did not seek clarification, users had to refine their prompts multiple times, increasing cognitive burden.

Finally, unresponsive answers to \textbf{prompt changes (F9, 9\%)} were particularly frustrating for users attempting to refine ChatGPT’s output. P29 saw no changes despite adjusting prompts, remarking ``\emph{I don't know if it's a problem with my prompts or with ChatGPT.}'' This inconsistency led to \textbf{user distrust}, as modifications did not always yield meaningful improvements.

\subsection{Failure Patterns in GPT-5.1 Replication}
We replicated the study using ChatGPT-5.1 \cite{OpenAI_2025} to examine whether previously identified failure categories persist in a newer model. While one-shot responses showed improvements, the same failure categories (F1–F9) continued to emerge during multi-turn task execution.

Interactions with GPT-5.1 continued to show \textbf{incomplete or incorrect responses (F1, F3, F7)} when users attempted step-by-step debugging. Although participants were able to generate a close replica of the website after sharing a screenshot of the webpage with minimal prompting techniques, follow-up edits frequently reintroduced missing details, overlooked required preconditions, or produced new code errors. 
P31 reported that the model’s responses were \textit{``getting worse the more [they] use it''}, and similarily, P32 stated, \textit{``The more I try to fix this, the worse it gets''}, mirroring the same patterns observed with legacy models when prompts accumulated complexity.

Furthermore, \textbf{cognitive overload and information mismanagement (F2, F4, F8)} remained issues. When users requested small adjustments, GPT-5.1 continued to attempt to regenerate the entire codebase, obscuring the specific changes. While in some cases code localization was improved, the inconsistencies in answer quality remained an issue for users. 
This unpredictability undermined users’ ability to anticipate prompt effectiveness, limiting the usefulness of prior prompt-engineering experience.
During development, P31 continuously attempts to render the webpage within the chat to identify if a bug was user-caused or GPT-generated with the \texttt{preview} feature. However, the code compilation only worked sometimes, and in other times the compilation either produced an error message, did not show up, and does not have multi-file compilation abilities. P31 noted, \emph{``The way [ChatGPT] generates the code is really difficult to read. It's not meant to be readable.''}.

Finally, failures related to context loss and redundant effort (F5, F6, F9) were still evident. As conversations lengthened, GPT-5.1 was prone to hallucination, sometimes generating entirely new codebases or stating that changes were made without any code differences. 

Despite added features and improvements in initial response quality, participants continued to express frustration when these changes failed to translate into more reliable multi-step code execution. As P32 summarized, \textit{``I’ve used ChatGPT for a long time, [GPT-5.1] added so many random features for the code to still not work.''}
These persistent issues limited GPT’s reliability in scenarios requiring sustained, iterative refinement, often contributing to user frustration and eventual disengagement.

\section{RQ2: Contributing Mechanisms of the Failures Observed} \label{causes}

ChatGPT aids SE, but failures stem from \textbf{user-contributed (UC)} and \textbf{ChatGPT-contributed (CC)} factors. In particular, task completion failures often stem from both user interactions mechanisms and limitations in ChatGPT’s design. Some failures arise because users struggle to communicate their intent effectively, while others occur due to ChatGPT’s inability to adapt to the complexities of real-world coding tasks. 
This section examines these issues, summarized in Tab.~\ref{tab:causes}.

While some task completion failures were initially triggered by user miscommunication, our analysis shows that many persisted or escalated due to ChatGPT-related limitations, including lost conversational context, poor code localization, and ignored instructions. These failures were not resolved through additional prompting or clarification.

\begin{table*} [t]
    \centering
    \footnotesize
        \caption{Contributions of Failure in user-LLM interactions, categorized into  user-contributed \textbf{(UC)} and  ChatGPT-contributed \textbf{(CC)} factors, with case counts across model versions. Overall, 49 unique cases were observed in sessions using GPT-4 and GPT-4o, and 23 cases were observed in the GPT-5.1 replication. P denotes participants.} 
    \begin{tabular} {@{} p{0.2cm} p{0.3cm} p{2.2cm} p{4.6cm} p{0.2cm} p{1cm} p{1cm} p{1.2cm} @{}}
        \toprule
\rotatebox{90}{\textbf{ }} & \textbf{ID} &
\textbf{Contributions of Failure} & \textbf{Description} & \textbf{\#P} & 
\textbf{GPT-4 (n, \%)} 
& \textbf{GPT-5.1 (n, \%)} 
& \textbf{Related Failures}\\\midrule

\multirow{6}{*}{\centering\rotatebox{90}{{\makecell[l]{\textbf{User Miscommunication}}}}} 
& UC1 & Vague or unspecific prompts & User questions were too general to be applied to the task & 13 & 22 (14\%) & 5 (12.5\%) & F1, F2, F3, F4, F5, F6 \\ 
\cmidrule{2-8}

& UC2 & Overly complex tasks & User delegated excessively complex tasks to ChatGPT, exceeding its capabilities. & 11 & 14 (9\%) & 2 (5\%) & F1, F2, F9 \\\cmidrule{2-8}

& UC3 & Skimming responses & Users skimmed responses, missing key details needed for resolution. & 4 & 5 (3\%) & 4 (10\%) & F1, F2, F3, F4, F6 \\
        \cmidrule{2-8}

& UC4 & Difficulty articulating needs & User lacked the necessary vocabulary or knowledge to describe their issue clearly. & 5 & 11 (7\%) & 4 (7.5\%) & F3, F4, F6, F9 \\ 
\cmidrule{2-8}

& UC5 & Inadvertent modifications & User unintentionally altered the code, introducing errors or unexpected behavior. & 2 & 2 (1\%) & 1 (2.5\%) & F5, F7 \\ 

\midrule
\multirow{7}{*}{\centering\rotatebox{90}{\makecell[l]{\textbf{ChatGPT's~~~Technical ~~~Limitations}}}} 
& CC1 & Ignored user expertise & ChatGPT responses exceeded the user's skill level. & 12 & 15 (10\%) & 3 (7.5\%) & F1, F2, F3, F4 \\ 
\cmidrule{2-8}

& CC2 & Weak file interaction support & ChatGPT failed to provide helpful guidance on handling or modifying files. & 10 & 16 (11\%) & 4 (10\%) & F1, F2, F3, F4, F5, F9 \\ 
\cmidrule{2-8}

& CC3 & Poor code localization & ChatGPT lacked code change indicators, requiring repeated clarifications. & 8 & 21 (14\%) & 2 (5\%) & F1, F3, F4, F5, F9 \\ 
\cmidrule{2-8}

& CC4 & No visual output support & ChatGPT could not generate or verify visual representations of its code output. & 2 & 4 (3\%) & 2 (5\%) & F6, F8 \\ 
\cmidrule{2-8}

& CC5 & Lack of code validation & ChatGPT produced incorrect or conflicting code without verifying correctness. & 6 & 9 (6\%) & 3 (7.5\%) & F5, F7, F9 \\ 
\cmidrule{2-8}

& CC6 & Lost conversation context & ChatGPT did not recall relevant past responses, forcing users to repeat information. & 4 & 18 (12\%) & 3 (7.5\%) & F1, F4, F5, F7 \\ 
\cmidrule{2-8}

& CC7 & Ignored prompt instructions & ChatGPT failed to follow explicit instructions in the user’s prompt. & 5 & 16 (10\%) & 12 (30\%) & F2, F7, F9 \\  
 
\bottomrule
    \end{tabular} 
    \label{tab:causes}
\end{table*}

\subsection{\textbf{User Miscommunication: When Expectations Exceed Clarity}} 
ChatGPT is only as effective as the prompts it receives, and many failures stemmed from how users framed their questions and processed the responses.

A common issue was \textbf{vague or underspecified prompts (UC1)}, where users failed to provide sufficient detail. For instance, P14 asked, ``\emph{How do I create a webpage in HTML?}'', which is too broad with no context about the desired features, design, or technologies. ChatGPT returned an extensive, generalized answer, which failed to meet the user’s specific needs, contributing to six types of failures \textbf{(F1-F6)}. This issue was particularly evident among users with limited technical expertise, who often expected ChatGPT to infer details rather than providing explicit instructions.

Compounding this, some users posed  \textbf{overly complex,multi-step tasks (UC2)}, expecting ChatGPT to generate fully functional solutions in a single interaction.  This often resulted in \textbf{F1 (Incomplete Answer)}, where ChatGPT missed key components, or  \textbf{F3 (Missing Preconditions)}, where it skipped setup steps like installing dependencies or linking files. In other cases, ChatGPT attempted to process everything at once, producing  \textbf{F2 (Overwhelming Answer)}, a lengthy, difficult-to-navigate response filled with unnecessary details. P27 explained ``\emph{Sometimes ChatGPT will give me complete garbage. But when its answer is done all at once, I do have to copy the entire thing.}''. Instead of helping users break tasks down logically, ChatGPT’s responses in these cases often left them struggling to extract useful information or refine their queries for more actionable results.

\begin{figure} [h]
    \includegraphics    [width=0.25\columnwidth]    {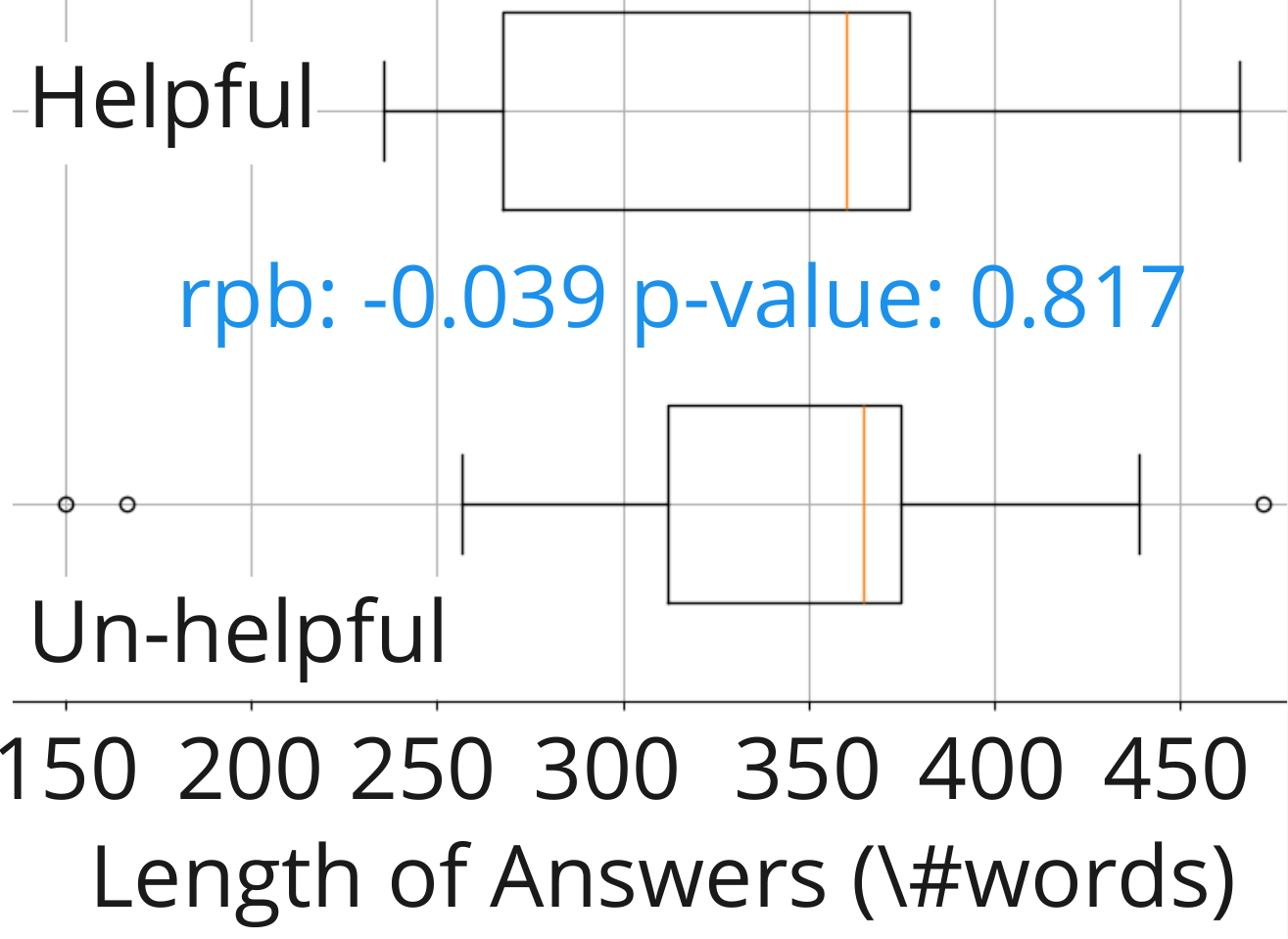}
    \caption{Boxplot of ChatGPT answer length, by helpfulness.}
    \Description{Boxplot showing the distribution of ChatGPT answer lengths across helpfulness ratings. Each box represents the interquartile range with a median line, and whiskers indicate the spread of values.}
    \label{length}
    \vspace{-1.5em}
\end{figure}

Our observations suggest users sometimes overlooked key explanations in ChatGPT’s responses \textbf{(UC3)}, skimming long outputs.
While one might assume that longer responses correlated with failures, statistical analysis, seen in Fig.~\ref{length}, showed no significant difference ($p = 0.817$) in response length between helpful and unhelpful cases. This suggests that failure was not due to verbosity but rather by how users processed the responses.

Another challenge was articulating queries effectively  \textbf{(UC4)}. Users unfamiliar with technical terminology struggled to describe their issues in a way that ChatGPT could interpret correctly. For example, when attempting to adjust CSS styles, some users asked vague or ambiguous questions, resulting in \textbf{irrelevant (F6) or overcomplicated (F4) answers}. Instead of simple inline styles, ChatGPT sometimes recommended unnecessarily complex solutions, assuming the user had expertise they did not possess.

In some cases, users \textbf{inadvertently modified ChatGPT’s responses (UC5)}, introducing errors that they later misattributed to ChatGPT. For instance, P21 manually renamed a CSS file without realizing that ChatGPT’s code was dependent on a specific filename, causing the webpage to break. This highlights a subtle but important issue: even when ChatGPT provides a functional solution, user modifications can introduce new points of failure.

\subsection{ChatGPT Adaptation Failures}
While user behavior played a key role in failures, ChatGPT’s design flaws also contributed significantly. One major issue was its \textbf{inconsistent adaptation to user expertise (CC1)}. Depending on the query phrasing, ChatGPT sometimes provided \textbf{overly advanced explanations} and, at other times, \textbf{overly simplistic ones}. P17, for example, received a complex response involving CSS \texttt{nth-child selectors}, while P21, who had a similar level of experience, was given a much simpler inline styling solution. This inconsistency made it difficult for users to predict how detailed or simplified ChatGPT’s responses would be.

Another recurring limitation was \textbf{ChatGPT’s inability to manage file interactions effectively (CC2)}. When responding to questions that required modifying multiple files, ChatGPT often renamed files arbitrarily (e.g., switching between \texttt{style.css} and  \texttt{styles.css} within the same conversation) or failed to indicate where new code should be placed. This lack of awareness contributed to \textbf{F1 (Incomplete Answers), F5 (Lacks Context), and F9 (Unresponsive to Prompt Changes)}, forcing users to manually debug inconsistencies in file references.

Similarly, ChatGPT \textbf{struggled with code localization (CC3)}, failing to indicate where a change should be made. Instead of inserting modifications in the correct locations, ChatGPT frequently generated entirely new sections of code, leading users to abandon efforts to integrate the suggested fixes. This resulted in \textbf{incomplete (F1) and overcomplicated (F4) solutions}, as users struggled to merge ChatGPT’s suggestions into their existing work.

For tasks requiring visual feedback, ChatGPT performed particularly poorly. Some users asked ChatGPT to generate previews of webpage layouts or UI elements \textbf{(CC4)}, only to receive misleading or incorrect visual representations. Since ChatGPT cannot render actual front-end designs, these failures \textbf{(F6: Irrelevant Answers, F8: Incorrect Image Generation)} misled users into believing that they had successfully implemented the correct styling.

A more subtle but critical issue was ChatGPT’s \textbf{lack of verification mechanisms (CC5)}. It often generated code that conflicted with its own explanations, leading to \textbf{erroneous (F7) or misleading (F5) outputs}. Users who trusted its responses faced execution errors due to unchecked syntax, logic, or dependencies.

Finally, in some cases, ChatGPT \textbf{ignored explicit instructions (CC7)}. Users who specified clear constraints or conditions in their prompts often found that ChatGPT deviated from their requests, instead generating responses based on its default assumptions. These failures \textbf{(F1: Incomplete Answer, F2: Overwhelming Answer, F7: Erroneous Answer, F9: Unresponsive to Prompt Changes)} demonstrate that even direct, well-formed queries do not guarantee compliance, adding an element of unpredictability to ChatGPT’s reliability.

\subsection{Contributing Conditions to Interaction Breakdowns in GPT-5.1}
Building on the observed failure categories (RQ1), this section analyzes the contributing conditions of task completion failures and examines whether it persists across model versions, distinguishing between user-driven interactional challenges and limitations intrinsic to ChatGPT’s design.
Results from the GPT-5.1 replication indicate that the interaction mechanisms underlying failures remained largely consistent with earlier model versions, despite improvements in one-shot response quality.

\textbf{User-related contributions (UC1–UC5)} persisted almost unchanged, demonstrating that user behavior patterns remained consistent despite improved model behaviors and indicating that improved model performance did not eliminate frictions during interaction. 
Participants continued to provide underspecified prompts (UC1), bundle multiple tasks together (UC2), skim explanations (UC3), use imprecise terminology (UC4), or unintentionally break working code (UC5). These behaviors produced the same downstream failures, particularly incomplete answers (F1), missing preconditions (F3), irrelevant responses (F6), and unresponsive updates (F9), even as model capabilities improve. This suggests that model updates alone are insufficient to mitigate interactional frictions that are rooted in co-adapted human–AI practices.

On the model side, \textbf{core system limitations also persisted} in GPT-5.1.
Although participants perceived fewer failures overall, foundational challenges related to adaptation, context management, and edit locality remained evident. GPT-5.1 continued to exhibit inconsistent calibration to where initial responses appeared stronger but subsequent turns reverted to oscillations between overly advanced and overly simplistic explanations, contributing to overcomplicated (F4) or incomplete outputs (F1).

Limitations in \textbf{multi-file context handling} (CC2) also persisted. When tasks required coordinated edits across files, GPT-5.1 frequently failed to maintain alignment between file names, locations, and changes, leading to context loss (F5) and prompt-insensitive behavior (F9). Similarly, the model’s tendency to regenerate large code blocks instead of localizing edits (CC3) remained, increasing cognitive load and hindering targeted correction during iterative debugging. Additional inconsistencies in preview rendering and internal coherence (CC4–CC7), although less frequent, continued to disrupt sustained refinement workflows.

Taken together, these findings indicate that while GPT-5.1 shifts the prevalence of certain failures, it does not fundamentally alter the mechanisms that give rise to breakdowns in human–LLM collaboration. Both user behaviors and model constraints co-evolve over time, and persistent interactional failures emerge from their interplay rather than from deficiencies on either side alone.

\section{RQ3: Mitigation Strategies} \label{mitigation}
In this section, we conceptualize mitigation as the set of adaptive strategies users employ to cope with repeated interactional failures during LLM-assisted development. These strategies reflect compensatory user effort rather than seamless collaboration, as users invest additional cognitive and temporal resources to recover from system limitations. Mitigation occurs after initial failures and often precedes either task completion through external means or disengagement from the LLM. Understanding mitigation therefore provides insight into how users attempt to sustain collaboration—and where that collaboration ultimately breaks down.

\subsection{Mitigation strategies with GPT-4 / GPT-4o}

\begin{table}[h]
\footnotesize
    \caption{Mitigation strategies observed during user–LLM interactions and their frequencies. The table reports the number and percentage of cases in the original study using legacy models (GPT-4 and GPT-4o) and in the GPT-5.1 replication. P denotes the number of participants who employed each strategy.}
    \centering
    \begin{tabular}{>{\raggedright\arraybackslash}
    p{0.23\linewidth}>{\raggedright\arraybackslash}
    p{0.43\linewidth}>{\raggedright\arraybackslash}
    p{0.01\linewidth}>{\raggedright\arraybackslash}
    p{0.08\linewidth}>{\raggedright\arraybackslash}
    p{0.08\linewidth}}
        \toprule
         \textbf{Mitigation }&  \textbf{Description} & 
         \textbf{\#P}&
      \textbf{GPT-4 (n, \%)} 
& \textbf{GPT-5.1 (n, \%)} 
        \\\midrule
          Further scaffold task& 
          Break the task into smaller subtasks for ChatGPT.
          & 19&32 (23\%) &12 (18\%)\\ \midrule
 Clarify the prompt & 
 Rephrase  the prompt  without adding new context.
 & 14&25 (18\%) &9 (14\%)\\ \midrule
  Switch task focus& 
  Abandon the request and shift focus.
  & 9&17 (12\%) &9 (14\%) \\ \midrule
Provide additional context&  
 Include extra code or details.
 & 11&16 (11.5\%) &13 (20\%) \\ \midrule
Manually debug ChatGPT’s output
 & 
 Copy, test, and refine ChatGPT’s code.
 & 16&21 (15\%) &14 (21\%) \\ \midrule
Point out ChatGPT’s mistakes
 &
 Identify and describe mistakes in ChatGPT’s response.
 & 14&21 (15\%) &5 (7.5\%) \\ \midrule

 Consult external sources& Leave ChatGPT and refer to a third-party resource.& 3&6 (4\%) &4 (6\%)
 \\\bottomrule
    \end{tabular}
    \label{tab:workarounds}
\end{table}

We first analyze mitigation strategies observed in interactions with legacy models (GPT-4 and GPT-4o), summarized in Tab.~\ref{tab:workarounds}. When ChatGPT’s responses were persistently unhelpful, users attempted a range of mitigation strategies, including prompt refinement, task scaffolding, and manual debugging. Despite these efforts, 17 of 26 participants ultimately abandoned ChatGPT during task completion, while seven proceeded by accepting unsatisfactory results.

\textbf{Iterating on Prompts: A Partial Fix at Best.} \textit{P: 4-22, 24-30.}
Rewording prompts was a go-to strategy, but its effectiveness depended on the type of failure. When ChatGPT’s response was factually \textbf{incorrect (F7) or lacked context (F5)}, prompt refinements occasionally steered it toward better answers. However, clarifications alone rarely addressed deeper reasoning flaws. ChatGPT might confidently generate different but still incorrect responses, showing that users needed more than just better phrasing to overcome systemic issues. Adding code snippets or external constraints improved some cases, but even experienced users found this unpredictable, as ChatGPT often failed to integrate new context cohesively across multiple turns.

\textbf{Recognizing ChatGPT’s Blind Spots: Debugging vs. Error Spotting.} \textit{P: 4, 5, 7, 9, 14, 15, 21, 25, 26, 28-30.}
Users who manually tested and debugged ChatGPT’s output were more successful in resolving certain types of failures—particularly \textbf{incomplete answers (F1), missing preconditions (F3), and erroneous responses (F7)}. Unlike re-prompting, which relied on ChatGPT self-correcting, debugging allowed users to identify gaps, correct logic, and iteratively refine the AI-generated code themselves.

\textbf{Task Restructuring: When Breaking It Down Still Fails.} \textit{P: 4-9, 14-19, 21, 22, 25-27, 29, 30.}
Scaffolding tasks into smaller steps often helped users obtain clearer responses and addressed failures such as \textbf{incomplete answers (F1), missing preconditions (F3), and overwhelming responses (F2)}. However, the assumption that smaller queries always improve results was not universally true—ChatGPT sometimes lost track of dependencies, leading to fragmented or inconsistent outputs. When users attempted to chain multiple responses together, the AI occasionally failed to maintain coherence across iterations, forcing them to manually re-integrate information.

\textbf{When Users Abandon ChatGPT on Task Completion.} \textit{P: 4, 15, 28.}
Some users abandoned ChatGPT for external sources (e.g., Google, StackOverflow, Reddit) when it struggled with vague or exploratory queries, leading to misleading (F9) or circular responses.
This led the users to recognize when AI was unproductive, reinforcing its value for well-defined tasks over open-ended exploration.
In these cases, users experienced continual failures, creating the moment of realization for users that their task was outside of ChatGPT's abilities.
For example, with P28, at 00:25:42 (Chat interaction count (QnA) 2), the prior ChatGPT's answers were incomplete (F1), and ignored key prompt details (F9).
In response, P28 stated, ``\emph{I don't trust [ChatGPT] that much when it comes to fixing the minor things. This is not a good fix.}'' 
After this response, P28 entered no further prompts.
At 00:28:00, following a pause of more than two minutes, P28 discontinued the use of ChatGPT and switched to manual editing. We coded this as an abandonment point because the participant provided an explicit verbal cue, stopped prompting altogether, and exhibited a sustained temporal gap that indicated disengagement rather than momentary frustration. This interpretation aligns with our triangulation method described in Sec.~\ref{factors}.

In other cases, abandonment stemmed from the user’s own uncertainty rather than ChatGPT’s explicit failures.
At 00:23:28 (QnA turn 5), after struggling to refine a description of Task C, P4 noted, ``\emph{I feel like this [task C] might take me down a rabbit hole. I can't really describe it to ChatGPT.}'' 
Following this statement, P4 paused for 7 seconds without entering another prompt and then shifted to implementing the solution manually for about  19 minutes. We coded this point as abandonment because the participant articulated an inability to progress with ChatGPT, ceased all subsequent prompting, and redirected their effort away from the tool—behavior consistent with our triangulation criteria for identifying disengagement (Sec.~\ref{factors}).
While ChatGPT may have been capable of providing high-quality answers that matched the users' needs, it was not able to bridge the knowledge gap sufficiently for the solution to be understood or applied.

In these cases, users found traditional search engines to be more helpful in bridging the knowledge gap. Professionals tended to consult external sources more and were more flexible in incorporating both ChatGPT answers and answers from Google. Participants commonly used Google for looking for more visual tasks than debugging, such as P4 referring to Google images for styling help, as they found the visual examples to be more helpful than ChatGPT's step-by-step solution. 
 Other resources included online tutorials, Reddit, and StackOverflow, where participants would compare the results with ChatGPT. 
    Participants noted that online tutorials were more engaging and skimmable, as responses were provided in increasing levels of difficulty and scaffolded consistently. 
    SDEs found that Reddit and StackOverflow had answers with details that targeted programmers, rather than ChatGPT's more generalized written responses.
Out of all participants, only two students switched to external sources, while three out of five SDEs consulted external sources. 

\subsection{Mitigation Strategies in GPT-5.1}
To examine whether mitigation strategies changed with newer models, we analyzed user behavior in the GPT-5.1 replication using the same coding framework. Participants employed the same mitigation strategies identified in legacy models, as shown in Tab.~\ref{mitigation}, indicating strong transferability of mitigation behaviors across model versions.

Participants employed similar mitigation strategies regardless of model updates. Despite improvements in GPT-5.1, users still required follow-up interactions to refine solutions. The model now frequently includes follow-up suggestions at the end of answers, which many participants used to scaffold tasks or switch focus. However, relying on these suggestions sometimes led to task derailment: for example, P31 spent over 20 minutes attempting to add animations, even though no such feature was required.

Participants (P31, P32, P33 and P36) continued to consult external sources, which often coincided with moments of abandonment. For instance, after following an online tutorial, P31 explained, ``\emph{I don't want to keep using [ChatGPT]. The [online tutorial] is more fun. It feels like I'm actually coding.}''
Similarly, P33 compared StackOverflow answers to ChatGPT, observing, ``\emph{The examples are helpful, it has fewer explanations, and the code's more skimmable when I'm reading.}''
As well, during their Google search, P33 relied on the search autocomplete function to better phrase their question to include more keywords. They noted that it helped them to better identify their question and helped externalize relevant vocabulary, improving their recall of relevant code. 
Overall, give-up points were consistent, resulting from a combination of a realization of ChatGPT's limits after repeated incomplete answers or unresponsive answers, consistent with our observations in prior GPT-4 versions.

\subsection{Synthesizing Failures, Causes, and Mitigation Strategies.}
\begin{figure}[h]
    \centering
    \includegraphics[width=0.8\columnwidth]{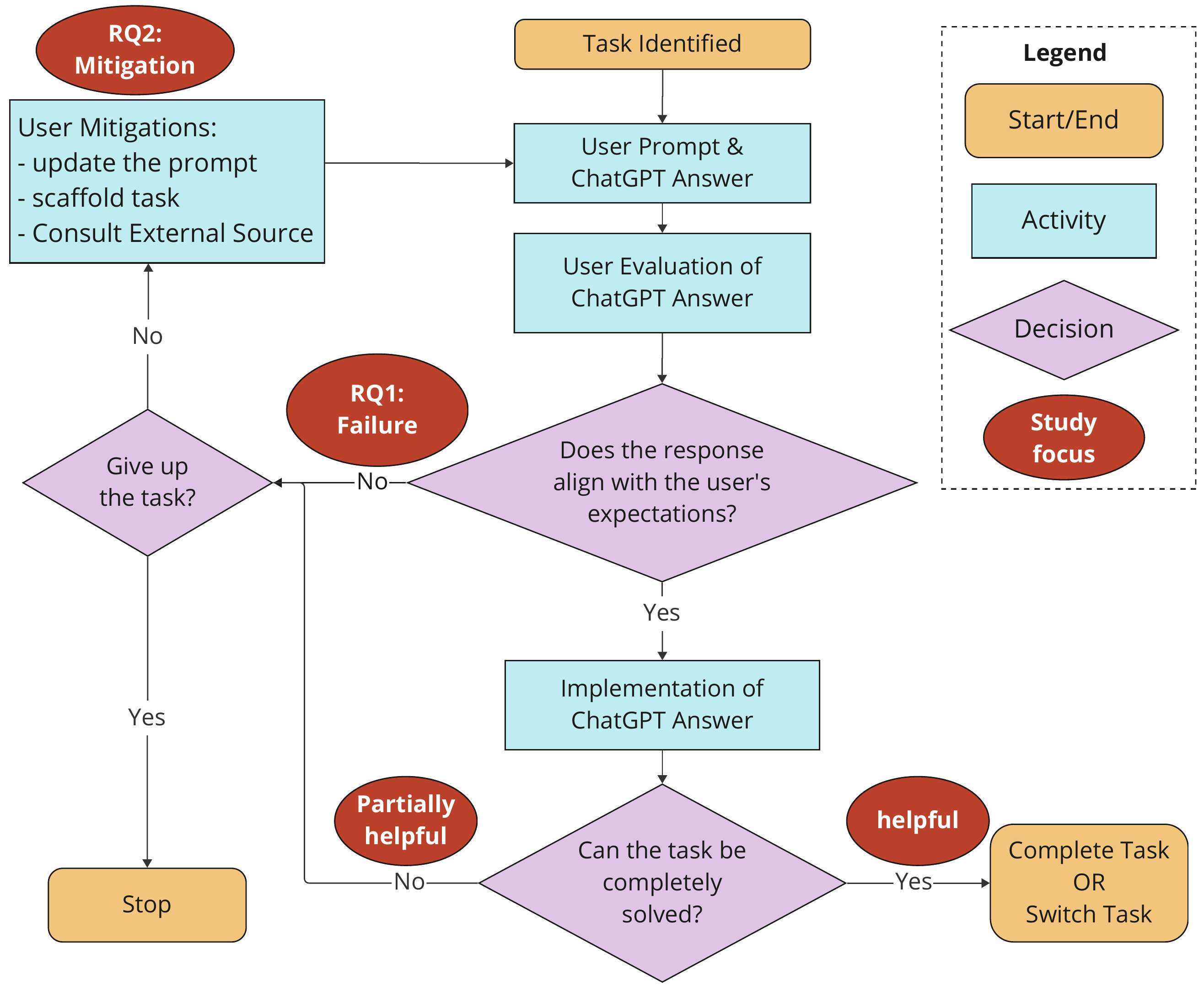}
    \caption{Workflow of user interaction with ChatGPT: focus on failure appearance and mitigation strategies. Unhelpful answers occur with failures.}
    \Description{Workflow of user steps when using ChatGPT for task completion.}
    \label{fig:workflow}
\end{figure}

To illustrate the connections between failures, their causes, and user mitigation strategies, we provide two key visualizations:

\begin{figure}[h]
    \includegraphics[width=0.5\columnwidth]
{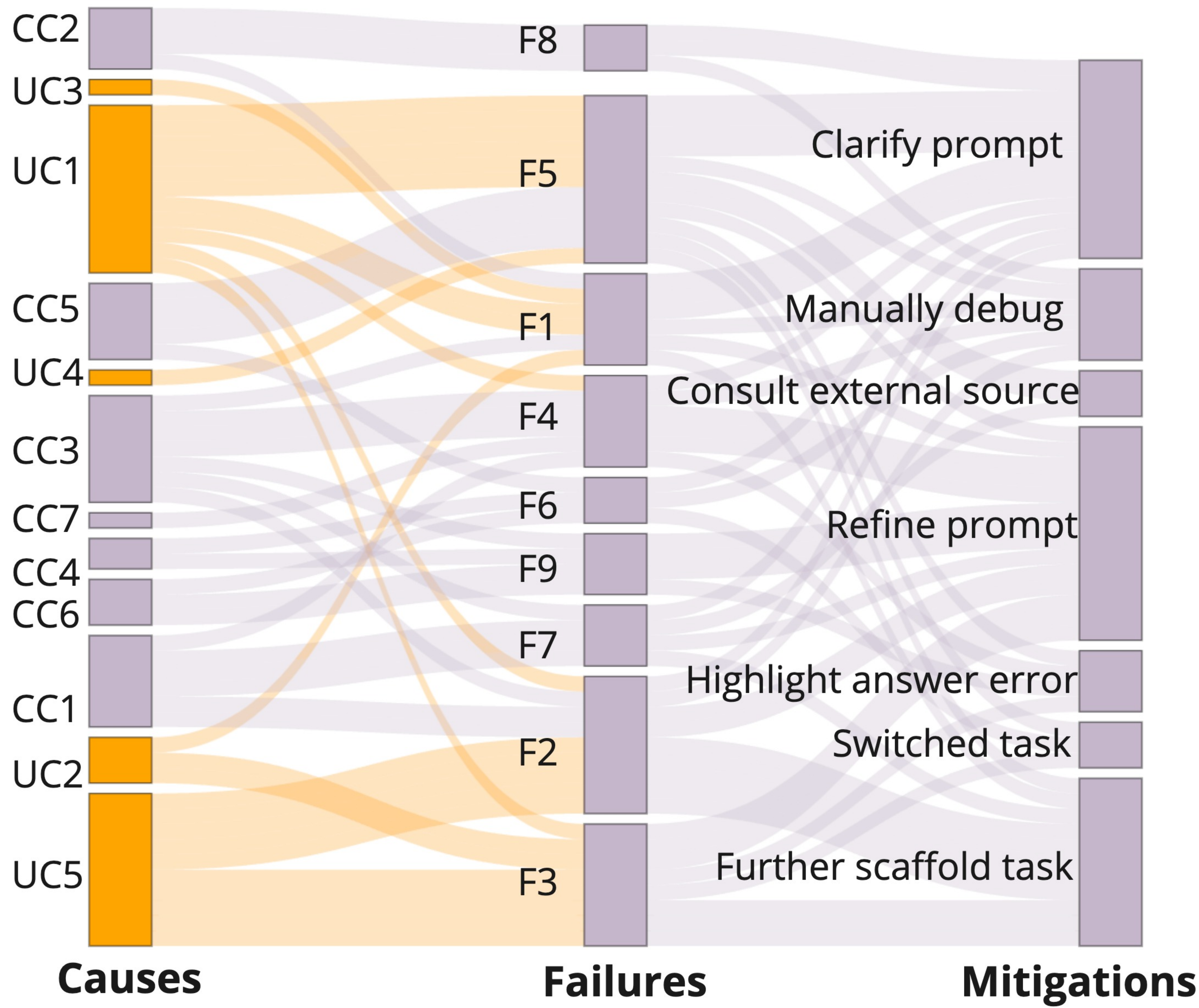}
    \caption{Connecting Causes \& Failures \& Mitigations.}
    \Description{Sankey diagram connecting contributing causes to associated failures and mitigations.}
    \label{fig:sankey}
\end{figure}
\begin{itemize} [leftmargin=1em] 
\item \textbf{Sankey diagram (Fig.~\ref{fig:sankey})} mapping failure causes (UC \& CC), observed failures (F1–F9), and mitigation strategies. It highlights recurring patterns, showing how \textbf{vague prompts (UC1)} often lead to \textbf{incomplete (F1)} or \textbf{irrelevant (F6)} responses, while \textbf{lack of verification (CC5)} causes \textbf{erroneous outputs (F7)}. Users mitigated failures by refining prompts or debugging, but persistent issues like ignored \textbf{modifications (F9)} often led to external help or task abandonment.
\item \textbf{User-ChatGPT workflow (Fig.~\ref{fig:workflow})} the iterative nature of human-LLM interaction, where users evaluate responses and refine prompts, seek external sources, or abandon tasks. The effectiveness of oversight depends on user experience, skilled users can navigate failures efficiently, while others may fall into a loop of ineffective refinements. This underscores the need for user adaptability, positioning LLMs as augmentative tools rather than autonomous problem solvers.
\end{itemize}

\section{RQ4\&5: User Perceptions, Task Abandonment, and Associated Factors}

In this section, we examine user perceptions of ChatGPT’s effectiveness (\textbf{RQ4}) and analyze factors associated with the likelihood of abandoning ChatGPT for task completion (\textbf{RQ5}). We first present qualitative findings from participant reflections on usability and confidence levels, followed by a quantitative analysis of abandonment tendencies and correlational results.

\subsection{RQ4: Reflection of ChatGPT Helpfulness}

\begin{figure}[h]
    \centering
    \includegraphics[width=0.9\columnwidth]{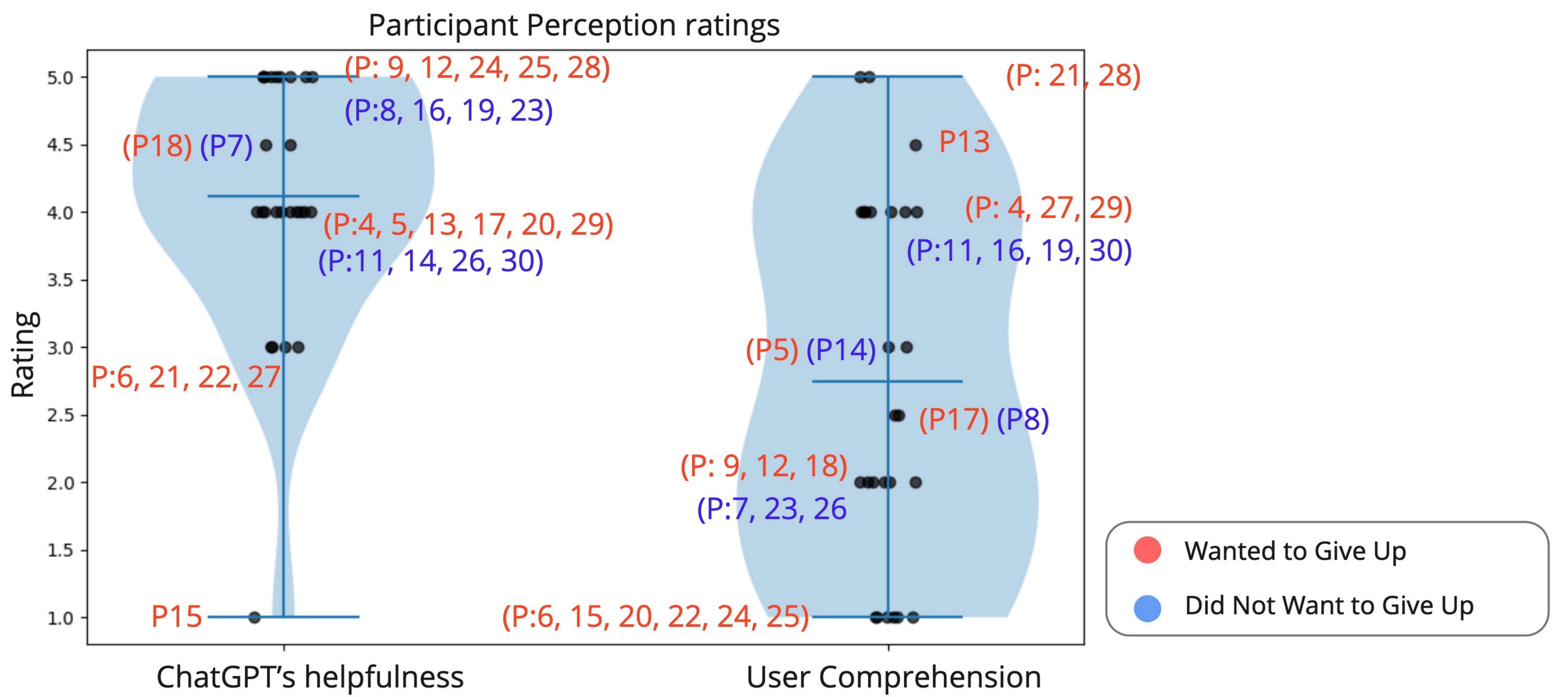}
    \caption{Participant's ratings on ChatGPT's helpfulness and code 
    comprehension support. Participants are listed and categorized based on whether they wanted to give up on using ChatGPT or not.}
    \Description{Violin graph of participants' ratings on ChatGPT, with each participant's rating marked.}
    \label{fig:survey-quant}
\end{figure}

Fig.~\ref{fig:survey-quant} shows that most participants rated ChatGPT’s helpfulness positively (3–5 on a Likert scale), although code readability ratings were more varied (1–3, with some rating 4 or 5). These results highlight user variability: P15's low rating reflected difficulties in code comprehension, negatively influencing their perception of ChatGPT, whereas P21, despite limited HTML experience, provided higher readability ratings, suggesting familiarity with coding structures enhances perceived helpfulness.

While ChatGPT provided valuable assistance in structured scenarios, many participants expressed frustration due to inconsistent responses, missing preconditions, and challenges with prompting refinement. These frustrations correspond to several identified failure categories, particularly incomplete answers (F1), overwhelming answers (F2), answers lacking context (F5), and irrelevant responses (F6).

\textbf{Professional developers frequently face unfamiliar tasks, making ChatGPT especially valuable.} All SDEs working reported regularly encountering unfamiliar code and tasks. P26 noted, \textit{``About 40–50\% of the work is new. A lot of the time, knowledge is new, and I need to go through documentation to continue my work. That’s when I rely on ChatGPT to help.}'' However, while ChatGPT effectively summarized code, it did not always facilitate deeper learning. For example, P28 reported difficulty grasping the generated code structure: \textit{``ChatGPT generated the code, but I still had no idea how the hierarchy worked. I hadn't learned that before.}'' This highlights a limitation—while assisting with immediate tasks, ChatGPT may not adequately support deeper understanding or knowledge transfer in new domains or languages.

\textbf{Need More Support for UI/UX.}
20 participants (P4, 5, 9, 12–15, 18, 20–30) expressed frustration with code localization, debugging, and tool interaction. These challenges align with previously identified problems in user-chat interactions and code complexity, and users also noted a desire for improved design, particularly in terms of UX.

P20, P22, and P28 criticized ChatGPT’s poor UX design skills, while effective prompting was a major challenge. P12 remarked, ``\textit{the only effort is put into prompting, not coding,}'' P24 struggled with precision, and P17 was unsure what to ask next. P28 stated, ``\textit{I'm not sure how ChatGPT perceives visuals.}'' These insights highlight that mastering prompting is essential, as difficulty formulating queries can hinder motivation and usability.

\textbf{Confidence in ChatGPT’s effectiveness varied widely.}
Regarding future use, five participants indicated they would continue using ChatGPT for similar tasks. Seven participants   preferred external sources such as online resources or documentation. Thirteen participants preferred manual coding or expert assistance. Confidence levels varied: six participants felt more confident about similar tasks in the future, ten participants would only feel more confident with continued use of ChatGPT, and four participants did not feel more confident, citing a lack of understanding and the need to know how to code independently.
Participants who persisted had the highest completeness scores; however, some who abandoned outscored others who quit despite wanting to continue, suggesting individual differences in satisfaction and effort perception.

\textbf{Users abandoned tasks when ChatGPT's perceived effort exceeded its benefits.} 
Although most participants rated ChatGPT as helpful overall (4/5), 17 of 26 users reported moments where they wanted to stop using it. As shown in Sec.~\ref{mitigation}, this often followed repeated failures, increasing prompting effort, or outputs that diverged from user expectations. In these situations, participants recalculated whether interacting with ChatGPT was still worthwhile relative to completing the task manually.

Several participants described abandoning ChatGPT after realizing that further refinement would be unproductive.
For example, P29   felt that repeated attempts were ``\textit{ wasting time on something that doesn't work.}''  
Others settled for suboptimal results because further prompting seemed unproductive; P11 described accepting a result that differed from expectations as ``\textit{good enough},'' while P18 and P27 reported that continued prompting had become repetitive and not worth the effort. These examples suggest that ChatGPT’s limitations can discourage further refinement once the user perceives diminishing returns.

Abandonment was also shaped by users’ broader expectations and professional norms.
Some users did not explicitly give up on ChatGPT but completed the tasks independently when they believed manual work would be more reliable or efficient.
SDEs, in particular, reflected that in real-world scenarios, they would likely have to manually continue editing the code without ChatGPT, as it would be less work than dealing with any potential errors that ChatGPT generates. 
As P27 summarised, ``\emph{Personally, anything that involves functional portions of the website, I prefer to do it myself, just because it could break down. If I'm at work, I'm not letting ChatGPT touch any of the functional parts of the project. It's just best practice.}''
Even if ChatGPT generated perfect code each time, SDEs suggested that real-world scenarios would still see the requirement of manual coding, consulting external sources, and prompting would become tedious as the task became smaller and more manageable. 
Students echoed similar sentiments when tasks became unfamiliar or cognitively demanding, often reverting to independent effort despite ChatGPT’s potential to generate helpful code.

Together, these behaviors highlight that abandonment is not merely a reaction to isolated errors. Instead, it reflects an emerging effort–benefit tradeoff in LLM-assisted development, where users continually assess whether the marginal value of further prompting justifies the cognitive and temporal investment required. This dynamic distinguishes LLM abandonment from traditional software tool disengagement, which typically occurs at adoption time rather than unfolding progressively across an interaction. 
These interpretations are grounded in triangulated interaction evidence, as described in our qualitative coding and analysis procedures (Sec.~\ref{qual-analysis}).

\subsection{RQ5: Correlation between Tendency to Abandon ChatGPT and Helpfulness.}
We use logistic regression to analyze the relationship between ChatGPT response helpfulness, completeness scores, prompt count, coding experience, and abandonment likelihood, testing the four hypotheses. These results complement user perceptions, providing deeper insights for improvement.  
As described in Sec. ~\ref{factors}, we modelled each participant’s coding process by subtask, constructing a vector representation that captures subtask segmentation, ChatGPT response implementation, and contextual factors.  

\subsubsection{Hypothesis Testing Result} \label{hypo_results}
To evaluate our hypotheses regarding the relationship between participants' tendency to abandon ChatGPT for task completion and factors such as 
the perceived helpfulness of ChatGPT responses  ($H_1$), 
project completeness ($H_2$), the number of prompts used ($H_3$), 
and participants' programming expertise ($H_4$),
we constructed a model to estimate factors influencing the likelihood of abandoning ChatGPT. The results are in Tab.~\ref{tab:glmresults}.

\begin{table}[t]
    \centering
    \footnotesize
    \caption{GLM estimation (binomial), DV: \textit{Abandon\_ChatGPT}}
    \vspace{-1em}
    \label{tab:glmresults}
    \begin{tabular}{l c c c}
        \toprule
        & \textbf{Model I} & \textbf{Model II} & \textbf{Model III} \\
        \midrule
        Partially Helpful\textsuperscript{\textdagger} 
            & 1.152 (0.530)\textsuperscript{**}& 1.218 (0.211)\textsuperscript{***}& \\
        Unhelpful 
            & 2.403 (0.547)\textsuperscript{***}& 2.382 (0.224)\textsuperscript{***}& 0.532 (0.434)\\
        Score 
            & 0.056  (0.406)& 0.051 (0.067)& 0.177 (0.067)\textsuperscript{**}\\
        \#Prompts 
            & -0.175 (0.018)\textsuperscript{***}& 0.279 (0.085)\textsuperscript{**}& -0.219 (0.010)\textsuperscript{**}\\
        \#CodingExp. [Yrs]
            & -0.730 (0.166)\textsuperscript{***}& & \\
        Student 
            & & 1.153 (0.199)\textsuperscript{***}& -0.933 (0.302)\textsuperscript{**}\\
        Student * \#Prompts 
            & & -0.510 (0.077) \textsuperscript{***}& \\
        Student * Unhelpful 
            & & & 1.070 (0.623)\\
        \midrule
        Observations & 302& 302& 262\\
        \bottomrule
    \end{tabular}
    \vspace{1ex}
    \begin{tabular}{l}
        \textsuperscript{*}\;p\,<\,0.05;\quad
        \textsuperscript{**}\;p\,<\,0.01;\quad
        \textsuperscript{***}\;p\,<\,0.001 \\
        \textsuperscript{\textdagger}\,Model~III removes \textit{helpfulness} due to few observations.
    \end{tabular}
        \vspace{-1.5em}
\end{table}

 \textbf{ ($H_1$) Relationship Between Helpfulness and Abandonment.}  
Regression analysis from Model I shows that, on average, the participants who received \texttt{unhelpful} responses from ChatGPT have around 11 times ($e^{2.403}$) higher odds of abandoning ChatGPT compared to those who received helpful responses. Similarly, 
\texttt{partially helpful} answers were associated with around three times ($e^{1.152}$) higher odds of giving up. \textbf{These results support $H_1$} and are consistent with qualitative reports in which participants expressed frustration over inaccurate or misleading responses. 
For example, P22 said, "I don't think the answer would be that helpful. I asked a question, and then I asked a different question, and the answers were the same."
This effect does not vary for either participant type (students and SDEs), as unhelpful answers were not significantly different for participant types seen from Model III, \texttt{Unhelpful: Student}.
This shows that while SDEs and students have different levels of prior knowledge, both groups ultimately abandoned the tool at similar rates when answers were unhelpful.

\textbf{($H_2$) Relationship Between Scores and Abandonment.}  
Results from \textbf{Model I} show that no significant relationship was observed for \texttt{scores}, leading us to \textbf{reject $H_2$}. We do not have enough evidence to conclude an association between the objective completeness of tasks and the odds of abandoning ChatGPT.

\textbf{($H_3$) Relationship Between Amount of Prompting and Abandonment.}  
As shown in \textbf{Model I}, each additional prompt reduces the odds of abandoning ChatGPT by 17\% ($e^{-0.175} - 1$), supporting $H_3$.
\textbf{Model II} further reveals that students are 40\% less likely ($1 - e^{-0.510}$) to abandon ChatGPT per extra prompt than SDEs, potentially indicating greater reliance among students.

\textbf{($H_4$) Coding Experience and Abandonment.}  
\textbf{Results also support $H_4$}, as greater \texttt{\#CodingExp.} was negatively correlated with the likelihood of abandoning ChatGPT. Participants with more coding experience may have been better able to adjust their prompts, interpret responses, or persist through initial difficulties. 
For example, P30, an SDE, noted, "If there was an issue with the answer, you can just rephrase the prompt," while P21, a student, said, "I don't know how else to prompt it," leading to abandoning ChatGPT.
\texttt{\#CodingExp.} is removed in \textbf{Models II-III} because it is highly collinear with participant type.
Notably, both groups still reported similar failure patterns and mitigation strategies.

\section{Discussion \& Implications}

\subsection{Model Updates and Persistent Interactional Failures}

Our study allowed us to observe developers working with LLM-generated code across multi-step tasks, reflective of real-world development processes. 
Unlike benchmark-driven or one-shot evaluations, this approach enables us to capture realistic, in-the-moment user interactions with AI systems.
The resulting dataset of user-prompts supports ongoing efforts to benchmark current LLM coding abilities ~\cite{xiao2024devgpt}, while also revealing interactional breakdowns that emerge only over sustained use.

Our observations spanned from GPT-4 and GPT-4o to GPT-5.1.
GPT-5.1 introduces several notable upgrades compared to earlier models, including the ability to generate near-complete website implementations using one-shot prompting, integrated code preview functionality that reduces reliance on image generation, more structured presentation of preconditions, and proactive follow-up suggestions or clarification questions intended to support task scaffolding. These changes reduce the amount of initial back-and-forth required to reach a workable baseline solution.

Despite multiple rounds of model improvements, user-AI interaction frictions continue to emerge. 
Rather than being eliminated by improved response quality alone, these frictions arose from the interactive nature of code refinement, where users repeatedly negotiate scope, context, and correctness with the model. 
In contrast to prior work that attributes breakdowns primarily to limited user expertise, these patterns were consistent across users with varying levels of experience, 
suggesting that such breakdowns are not solely attributable to insufficient user expertise ~\cite{kruse2024can}.

Our observations further showed that interaction frictions frequently arise during iterative attempts to improve generated code accuracy. 
While GPT-5.1 often produces stronger initial responses, follow-up turns remain unstable: minor edit requests frequently trigger partial rewrites, reintroduction of omitted details, or inconsistent refinements. 
Few-shot prompting and iterative clarification do not reliably stabilize these follow-up interactions, indicating that improvements in first-turn performance do not translate into more dependable multi-turn collaboration.

ChatGPT often overwrote large portions of existing code where users requested minor edits, introducing hallucination and context loss. 
These behaviors increased cognitive load and accelerated erosion of user trust, contributing to early abandonment, echoing prior findings on trust degradation in AI-assisted systems~\cite{brown2024identifying}, and persistent concerns around validation and reliability~\cite{tang2024study}. 

Some GPT-5.1 upgrades also introduce new pathways for breakdown. Code preview functionality, while useful for inspecting isolated outputs, remains limited to single files and does not support coordinated multi-file reasoning, leaving core context-management challenges unresolved. Similarly, automatically generated follow-up suggestions are frequently irrelevant to the task at hand and can lead users toward unnecessary features or extended “rabbit holes,” increasing time cost and confusion rather than facilitating progress.

When users encountered repeated failures, such as unintended errors in code changes or accumulating errors, frustration rapidly escalated, leading them to disengage from the tool. 
Here, abandonment emerged not as a binary decision but as a process shaped by accumulating interactional breakdowns.
This aligns with established theories in HCI that link cognitive overload and loss of situational awareness to task abandonment
~\cite{xia2018Measuring,sillito2006questions}.

Taken together, these findings suggest that while newer models substantially improve isolated response quality, they do not fundamentally alter the interactional mechanisms that give rise to failure during sustained use. Persistent limitations in context retention, edit locality, adaptation to user expertise, and output verification interact with user behaviors over time, producing cascading breakdowns that improvements in one-shot generation alone cannot resolve.

\subsection{Time Wasted during User-LLM Interaction}

We observed that users had difficulty skimming and chunking the code generated, as ChatGPT would generate line by line, as opposed to by sections, which is more natural to programmers for reading ~\cite{Prat_Madhyastha_Mottarella_Kuo_2020}.

This makes it harder for developers to quickly identify the errors, and to reduce mental load, all developers validated the code by manual execution, rather than attempting to parse through it ~\cite{Gentner_2014}. 
When developers have to switch between coding and debugging to prompt ChatGPT, they face interruptions to their focus, which can exacerbate the task workload, leading to disengagement ~\cite{russo2024navigating}. 

\subsubsection{Prompting Rabbit-Hole}
\label{sec:Prompting Rabbit-Hole}
Similar to debugging rabbit-holes, we observe users also engage in ``prompting rabbit-holes'', underestimating the necessary effort ~\cite{vaithilingam_2022}.  While frequent prompting generally reduced task abandonment ($H_3$), ineffective prompts increased abandonment rates significantly ($H_1$). 
SDE P29 and P14 had more prompts than others but differed in persistence. P29 gave up after four failed attempts and found a useful tutorial online. P14, despite 10 failures, blamed themselves and did not seek alternatives. This highlights how ChatGPT failed to guide P14 toward effective SE strategies.
To mitigate this, ChatGPT should detect unproductive prompting cycles and proactively provide clarification or guidance. Users should also compare LLM responses with external resources to assess their usefulness objectively. Our findings underscore the importance of AI systems identifying user struggles early and offering alternative strategies to avoid frustration.

\subsubsection{Overreliance}
Users frequently over-relied on ChatGPT, replacing entire codebases with each response instead of integrating incrementally, resulting in errors. This aligns with research showing uncritical AI acceptance can reduce comprehension and problem-solving skills \cite{passi2022overreliance}. While greater coding experience correlated with lower abandonment ($H_4$), SDEs abandoned tasks more quickly than students as prompts increased. Students persisted longer, often ignoring limitations, whereas SDEs switched sooner to alternative solutions. Users should integrate ChatGPT’s responses incrementally rather than replacing entire codebases, reducing errors and improving comprehension.

\subsection{Implications for SE researchers}
A key research area is mitigating user over-reliance on iterative prompting. 
Our findings provide empirical evidence that constructive ChatGPT interactions support sustained user engagement in coding tasks, reinforcing prior research on positive perceptions and persistence ~\cite{shah2023continuance, bosch2017affective}. 
Future work should identify unproductive prompting patterns and develop frameworks promoting efficiency and focus, such as periodic reflection prompts or automated summaries. Additionally, research into AI-driven engagement management could produce intelligent agents that proactively guide users toward task completion. 

Another important direction for future work is to understand how successful user-LLM interactions emerge through multiple interacting factors. 
Our construct of helpfulness reflects interactional progress rather than isolated response correctness, consistent with how developers experience AI-assisted workflows in practice.
Because response correctness alone is insufficient as a primary evaluation metric, future work should adopt multidimensional constructs of helpfulness that incorporate user-side and contextual factors when assessing AI systems.

\textbf{Enhancing Educational Value.}
Participants preferred traditional online resources over ChatGPT potentially due to deeper, more structured learning ~\cite{perkins_1989, denny_2024, okonkwo_2021}. Participants refined strategies like clarifying prompts and adding context, but frustration sometimes led to abandonment, highlighting missed opportunities for AI in SE learning. 
Improving ChatGPT’s educational value requires better guidance on prompt structuring, enhanced context retention, and quick error identification tools.

\subsection{Recommendations for LLM Designers}

\textbf{Balancing AI Predictions and User Judgment.}
Agrawal et al. \cite{agrawal_2022} define AI decoupling as replacing human predictions with AI-generated ones, shifting users to judgment-based evaluation. This occurs when users assess ChatGPT’s code: ChatGPT predicts a likely solution, and users judge its correctness. While this separation can broaden perspectives, it also demands stronger user judgment skills. Future LLM designs should emphasize critical evaluation over blind acceptance to enhance decision-making in code generation.

\textbf{Adapting Responses to User Expertise.}
ChatGPT influences user outcomes in programming and should adapt responses based on user expertise. Our findings show ChatGPT’s primary failure stems from not considering user expertise (CC1). Improvements should include adaptive responses and knowledge overviews to prevent overwhelming users. To resolve long, unstructured answers (F2), ChatGPT should scaffold code and highlight key details.

\textbf{Assessing Model Advancements in User Behavior.}
Our study spans the transition from GPT-4 to GPT-5.1. 
This setup enables us to assess whether model advancements influence user behavior and problem-solving strategies beyond improvements in output accuracy. Interestingly, despite GPT’s enhancements in software development accuracy, all models exhibited similar failures and mitigation effectiveness. This suggests that improving response correctness alone is insufficient, future LLM designs should also focus on guiding users through problem-solving processes and fostering adaptive learning strategies. 

\section{Conclusion}
We examined user-ChatGPT interactions in a SE context, identifying  failures and their causes. Both users and ChatGPT contribute to issues like incomplete and overwhelming answers. By pinpointing key failures and strategies, this research offers insights for enhancing LLM design and functionality to better align with problem-solving practices and user expectations. Future designs of LLMs should address the root causes of failures to better support users.
Future work should address these challenges and explore ways to optimize the integration of LLMs into real-world programming workflows.

\section*{Acknowledgement}
Zhou and Tie’s work was supported by the Natural Sciences and Engineering Research Council of Canada (NSERC), RGPIN2021-03538.
Ahmed’s work was supported by the School of Cities at the University of Toronto and by a Natural Sciences and Engineering Research Council of Canada (NSERC) Discovery Grant.

\newpage
\bibliographystyle{ACM-Reference-Format}
\bibliography{main}

\end{document}